\documentclass[twoside,twocolumn,9pt]{article}
\usepackage{extsizes}
\usepackage[super,sort&compress,comma]{natbib} 
\usepackage[version=3]{mhchem}
\usepackage[left=1.5cm, right=1.5cm, top=1.785cm, bottom=2.0cm]{geometry}
\usepackage{balance}
\usepackage{mathptmx}
\usepackage{afterpage}
\usepackage{placeins}
\usepackage{sectsty}
\usepackage{graphicx} 
\usepackage{lastpage}
\usepackage[format=plain,justification=justified,singlelinecheck=false,font={stretch=1.125,small,sf},labelfont=bf,labelsep=space]{caption}
\usepackage{float}
\usepackage{fancyhdr}
\usepackage{fnpos}
\usepackage[english]{babel}
\addto{\captionsenglish}{%
  \renewcommand{\refname}{Notes and references}
}
\usepackage{array}
\usepackage{droidsans}
\usepackage{charter}
\usepackage[T1]{fontenc}
\usepackage[usenames,dvipsnames]{xcolor}
\usepackage{setspace}
\usepackage[compact]{titlesec}
\usepackage{hyperref}

\usepackage{epstopdf}

\definecolor{cream}{RGB}{222,217,201}

\begin{document}

\pagestyle{fancy}
\thispagestyle{plain}
\fancypagestyle{plain}{
\renewcommand{\headrulewidth}{0pt}
}

\makeFNbottom
\makeatletter
\renewcommand\LARGE{\@setfontsize\LARGE{15pt}{17}}
\renewcommand\Large{\@setfontsize\Large{12pt}{14}}
\renewcommand\large{\@setfontsize\large{10pt}{12}}
\renewcommand\footnotesize{\@setfontsize\footnotesize{7pt}{10}}
\makeatother

\renewcommand{\thefootnote}{\fnsymbol{footnote}}
\renewcommand\footnoterule{\vspace*{1pt}%
\color{cream}\hrule width 3.5in height 0.4pt \color{black}\vspace*{5pt}} 
\setcounter{secnumdepth}{5}

\makeatletter 
\renewcommand\@biblabel[1]{#1}            
\renewcommand\@makefntext[1]%
{\noindent\makebox[0pt][r]{\@thefnmark\,}#1}
\makeatother 
\renewcommand{\figurename}{\small{Fig.}~}
\sectionfont{\sffamily\Large}
\subsectionfont{\normalsize}
\subsubsectionfont{\bf}
\setstretch{1.125} 
\setlength{\skip\footins}{0.8cm}
\setlength{\footnotesep}{0.25cm}
\setlength{\jot}{10pt}
\titlespacing*{\section}{0pt}{4pt}{4pt}
\titlespacing*{\subsection}{0pt}{15pt}{1pt}

\fancyfoot{}
\fancyfoot[LO,RE]{\vspace{-7.1pt}\includegraphics[height=9pt]{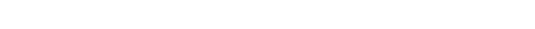}}
\fancyfoot[CO]{\vspace{-7.1pt}\hspace{13.2cm}\includegraphics{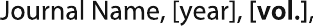}}
\fancyfoot[CE]{\vspace{-7.2pt}\hspace{-14.2cm}\includegraphics{head_foot/RF}}
\fancyfoot[RO]{\footnotesize{\sffamily{1--\pageref{LastPage} ~\textbar  \hspace{2pt}\thepage}}}
\fancyfoot[LE]{\footnotesize{\sffamily{\thepage~\textbar\hspace{3.45cm} 1--\pageref{LastPage}}}}
\fancyhead{}
\renewcommand{\headrulewidth}{0pt} 
\renewcommand{\footrulewidth}{0pt}
\setlength{\arrayrulewidth}{1pt}
\setlength{\columnsep}{6.5mm}
\setlength\bibsep{1pt}

\makeatletter 
\newlength{\figrulesep} 
\setlength{\figrulesep}{0.5\textfloatsep} 

\newcommand{\topfigrule}{\vspace*{-1pt}%
\noindent{\color{cream}\rule[-\figrulesep]{\columnwidth}{1.5pt}} }

\newcommand{\botfigrule}{\vspace*{-2pt}%
\noindent{\color{cream}\rule[\figrulesep]{\columnwidth}{1.5pt}} }

\newcommand{\dblfigrule}{\vspace*{-1pt}%
\noindent{\color{cream}\rule[-\figrulesep]{\textwidth}{1.5pt}} }

\makeatother

\twocolumn[
  \begin{@twocolumnfalse}
{\includegraphics[height=30pt]{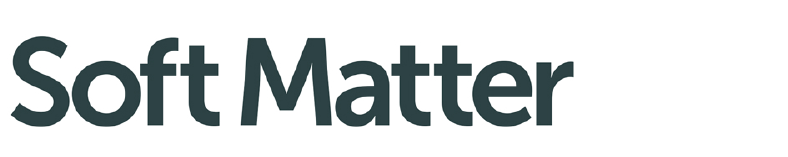}\hfill\raisebox{0pt}[0pt][0pt]{\includegraphics[height=55pt]{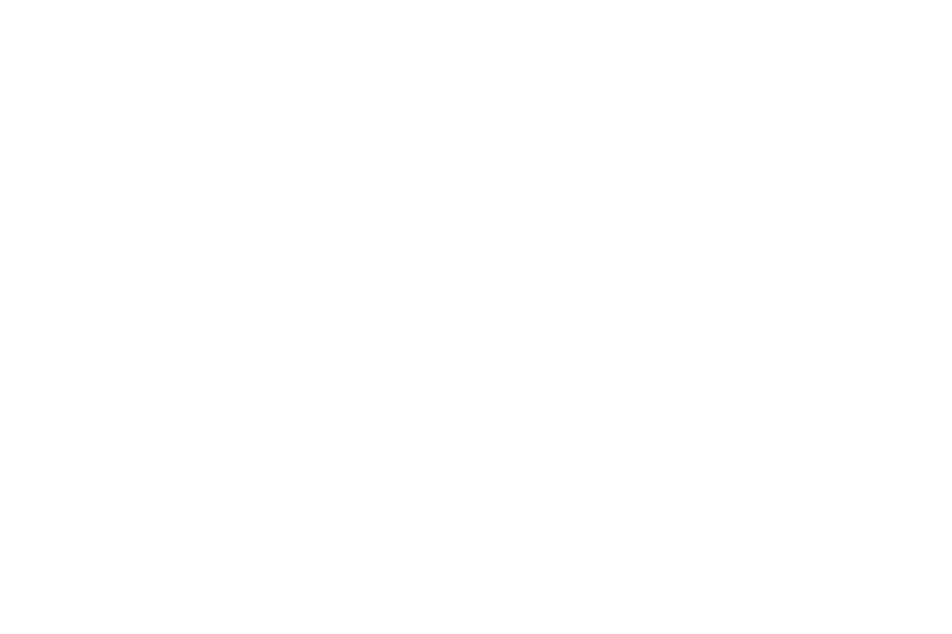}}\\[1ex]
\includegraphics[width=18.5cm]{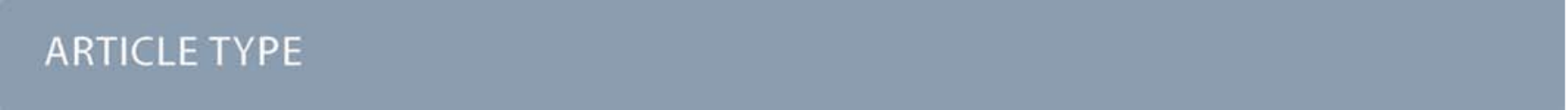}}\par
\vspace{1em}
\sffamily
\begin{tabular}{m{4.5cm} p{13.5cm} }

\includegraphics{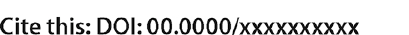} & \noindent\LARGE{\textbf{Amorphous Entangled Active Matter$^\dag$}} \\
\vspace{0.3cm} & \vspace{0.3cm} \\

 & \noindent\large{William Savoie,$^{\ddag}$\textit{$^{a}$} Harry Tuazon,$^{\ddag}$\textit{$^{b}$} M. Saad Bhamla,\textit{$^{b}$} and Daniel I. Goldman$^{\ast}$\textit{$^{a}$}} \\

\includegraphics{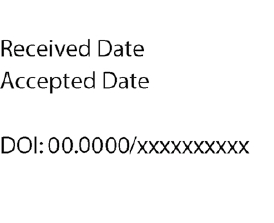} & \noindent\normalsize{The design of amorphous entangled systems, specifically from soft and active materials, has the potential to open exciting new classes of active, shape-shifting, and task-capable `smart’ materials. However, the global emergent mechanics that arises from the local interactions of individual particles are not well understood. In this study, we examine the emergent properties of amorphous entangled systems in three different examples: an in-silico ``smarticle'' collection, its robophysical chain, and living entangled aggregate of worm blobs (\textit{L. variegatus}). In simulations, we examine how material properties change for a collective composed of dynamic three-link robots. We compare three methods of controlling entanglement in a collective: externally oscillations, shape-changes, and internal oscillations. We find that large-amplitude changes of the particle's shape using the shape-change procedure produced the highest average number of entanglements, with respect to the aspect ratio ($l/w$), improving the tensile strength of the collective. We demonstrate application of these simulations in two experimental systems: robotic chains and entangled worm blobs. In the robophysical models, we find emergent auxeticity behavior upon straining the confined collective. And finally, we show how the individual worm activity in a blob can be controlled through the ambient dissolved oxygen in water, leading to complex emergent properties of the living entangled collective, such as solid-like entanglement and tumbling. Taken together, our work reveals principles by which future shape-modulating, potentially soft robotic systems may dynamically alter their material properties, advancing our understanding of living entangled materials, while inspiring new classes of synthetic emergent super-materials.} \\

\end{tabular}

 \end{@twocolumnfalse} \vspace{0.6cm}

  ]

\renewcommand*\rmdefault{bch}\normalfont\upshape
\rmfamily
\section*{}
\vspace{-1cm}


\footnotetext{\textit{$^{a}$School of Physics, Georgia Institute of Technology, Atlanta, GA 30318}}
\footnotetext{\textit{$^{b}$School of Chemical and Biomolecular Engineering, Georgia Institute of Technology, Atlanta, GA 30318 E-mail: daniel.goldman@physics.gatech.edu}}

\footnotetext{\dag~Electronic Supplementary Information (ESI) available: [details of any supplementary information available should be included here]. See DOI: xx.xxxxx/xxxxxxxx/}

\footnotetext{\ddag~These authors contributed equally to this work.}



\begin{figure}[h!]
\centering
    \includegraphics[width=\hsize]{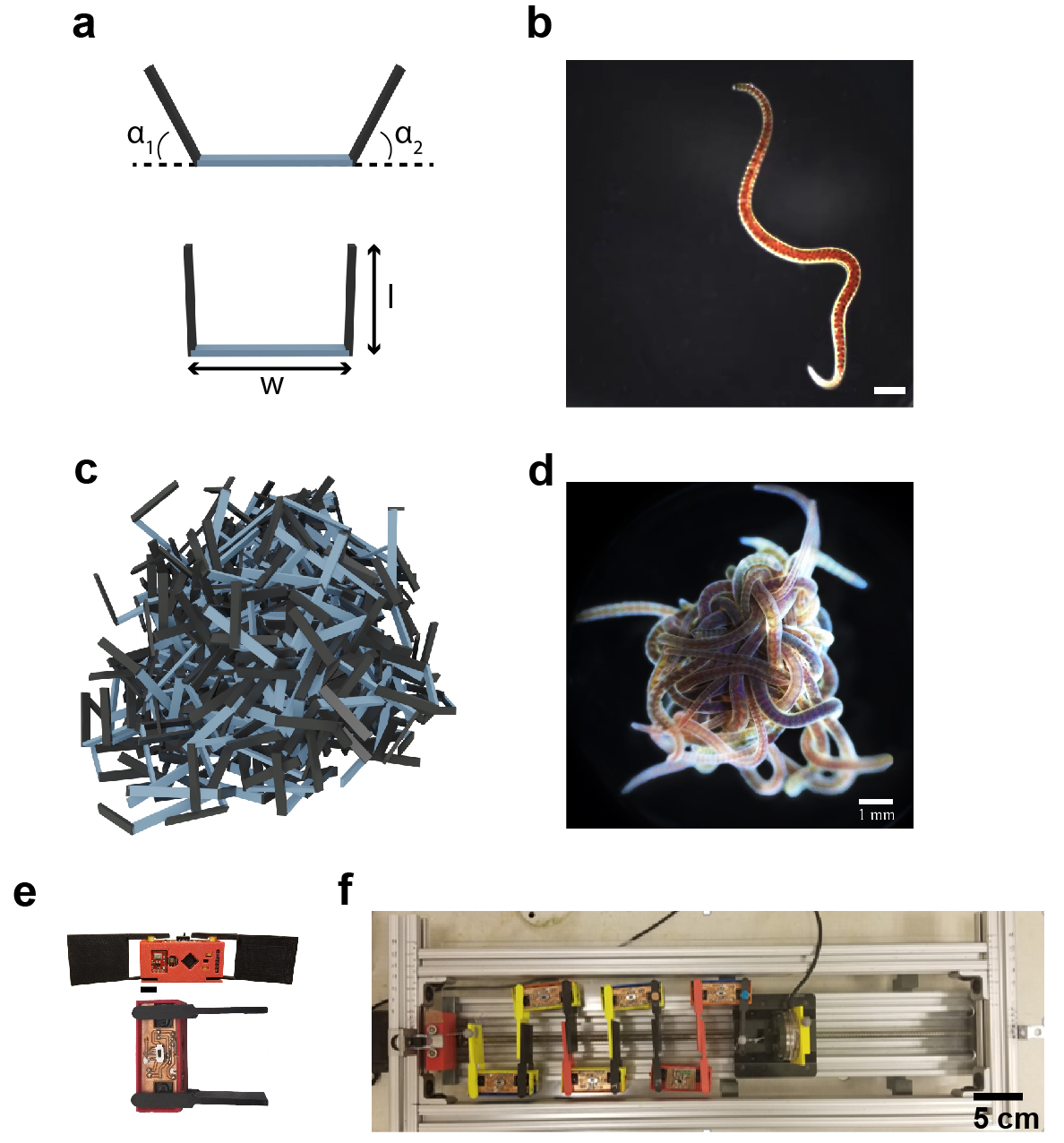}
    \caption[Smarticles in simulation and experiment] {\textbf{Smarticles in simulation, biological, and robophysical experiments.} (a) The coordinate system and size designations used for the smarticles. (b) An individual California blackworm (\textit{Lumbriculus variegatus}). (c) An entangled pile of smarticles in simulation. (d) A group of 20 individual blackworms forming a physically-entangled blob. (e) Front view of robotic smarticle used in experiment. Scale bar is 1 cm. (f) Smarticles in experiment.}
    \label{fig:simAndExpPics}
\end{figure}

\section{Introduction}


Physically entangled active matter is an emerging area of both living and man-made systems, where mechano-functionality of the collective emerges through physical interaction of individual elements. These are different from flocks of birds or swarms of fish that are not physically connected into an amorphous emergent material~\cite{2016Hu}. Biological examples includes cellular slime molds, insect assemblages and worm blobs, where the dynamics of the individual agent (cell, ant or worm) contributes to emergent functionality of the collective (slugs, rafts or blobs)~\cite{2012Reid,Mlot2011,2021Ozkan-Aydin}. Synthetic or biohybrid systems include xenobots~\cite{2021blackiston,2020kriegman}, acto-myosin assemblies~\cite{2014Dumont,1997Ndelec,2001Surrey}, granular materials~\cite{Gravish2012}, stochastic robot collectives~\cite{2021boudet,2019Li}, and super-smarticles~\cite{2019Savoie}. In contrast to biological counter-parts, robotic active matter systems are limited to spheroidal elements (low-aspect ratio), low density, non-interacting particles, or weakly actuating elements. 
Thus, P.W. Anderson’s famous observation that ‘More is different’ remains true, and foundational understanding through both theory and experiments that can guide bottom-up mechano-functional entangled active matter systems remains tantalizingly elusive~\cite{1980Anderson}.      



In soft condensed matter systems composed of particle ensembles, it is known that particle shape can influence rheological and structural properties like viscosity~\cite{Hsueh2010,Brown2011,Egres2005}, yield-stress~\cite{Brown2011,Kramb2011,murphy2019,nelson2019}, packing density~\cite{Man2005,Desmond2006,Miskin2014,Trepanier2010,Blouwolff2006}, and packing stiffness~\cite{miskin2013}. In granular media for example, there has been some research in the evolution of the macroscopic state dynamics. Most research was performed with external loading~\cite{Gravish2012,murphy2019}; however, less is known about how structural properties evolve as particle shape evolves, despite such transitions existing in nature (as in ant systems, in cytoskeleton and f-actin, in polymer rheology, and in aquatic worm blobs)~\cite{Gardel2003,foster2014,Mlot2012,mcleish2008,lieleg2009, 2021Ozkan-Aydin, 2021Nguyen, 2020aDeblais, 2020bDeblais}. In this work, we focus on an active matter system composed of shape-changing particles, where as we demonstrate below, rich and non-intuitive material physics can be uncovered. How macroscopic material properties are affected by the microscopic movements of its constituents is under-studied\cite{henein2007}. Moreover, for many-agent systems where elements can actuate, much of the research is either limited to planar systems\cite{garcimartin2015,zheng2009}, or are low density or non-interacting three-dimensional systems \cite{michael2006,kelley2013}. Yet as both computational units and actuating elements become smaller and cheaper, smaller and more highly interactive robotic systems may become more commonplace.


In this paper, we seek to discover and understand the independent formation of solid or semi-solid structures by an
open-loop dynamic collective. Using simulation, we examine how material properties change for a collective composed of dynamic particles whose form (a three-linked two degree-of-freedom shape) is based on the robotic particle called smarticles~\cite{Savoie2019,Savoie2018}. By implementing different shape-altering procedures, and probing the final structures, we gain insight into possible material properties attainable for future real-world robotic systems. Using stress-strain tests on an entangled chain of smarticles, we discover various properties related to their emergent auxeticity. Finally, we compare a few of these properties to biological experiment using California blackworms (\textit{Lumbriculus variegatus)}, whose collectives demonstrate a non-trivial macroscopic rheological analogue with features and properties from microscopic entanglement~\cite{2022Tuazon, 2021Ozkan-Aydin, 2021Nguyen}.

\section{Materials and methods}
To test the material properties of non-convex granular aggregates, we constructed a three-link robot in simulation Fig.~\ref{fig:simAndExpPics}(a,c) in robophysical experiments Fig.~\ref{fig:simAndExpPics}(e,f), and compared it to the entangled collective behavior demonstrated by blackworms Fig.~\ref{fig:simAndExpPics}(b,d).

\subsection{Multibody simulation of smarticles}
\label{sec:smartSim}
We developed a multibody simulated model of many staple-shaped three-link particles (henceforth called ``smarticles''), Fig.~\ref{fig:simAndExpPics}(a), capable of actuating their outer links. The smarticle's form was inspired by Purcell's three-link swimmer~\cite{Purcell1977,hatton2013} and the size was inspired by previous work from Gravish et al.~\cite{Gravish2012}. Two outer links (or barbs) are connected to a middle segment by rotational actuators, making the smarticle a planar system with two degrees-of-freedom. The middle link is of width $w=1.17$ cm and the barbs are of length $l=[0-1.1]w$ (all simulation parameters are shown in Tab.~\ref{tab:simPars}). The multibody simulation was implemented using ProjectChrono~\cite{ChronoWebsite,Tasora2015}, an open-source physics dynamics engine. 

Since smarticles can actuate their barbs to specific positions, each motor is velocity-controlled with a maximum allowable torque. Smarticle barbs are controlled to move with an angular velocity of $\omega = 6$ rad$/$s. The maximum allowable torque, $\tau_{max}$, was set to the torque required to lift a mass equivalent to a smarticle of $l/w=0.7$, a distance $w$ away from the actuator's axis. For $\tau_i \geq \tau_{max}$, the $i$th barb's movement was halted until the first time step where the condition $\tau_i<\tau_{max}$ was satisfied.

\begin{figure}[h!]
\centering
    \includegraphics[width=\hsize]{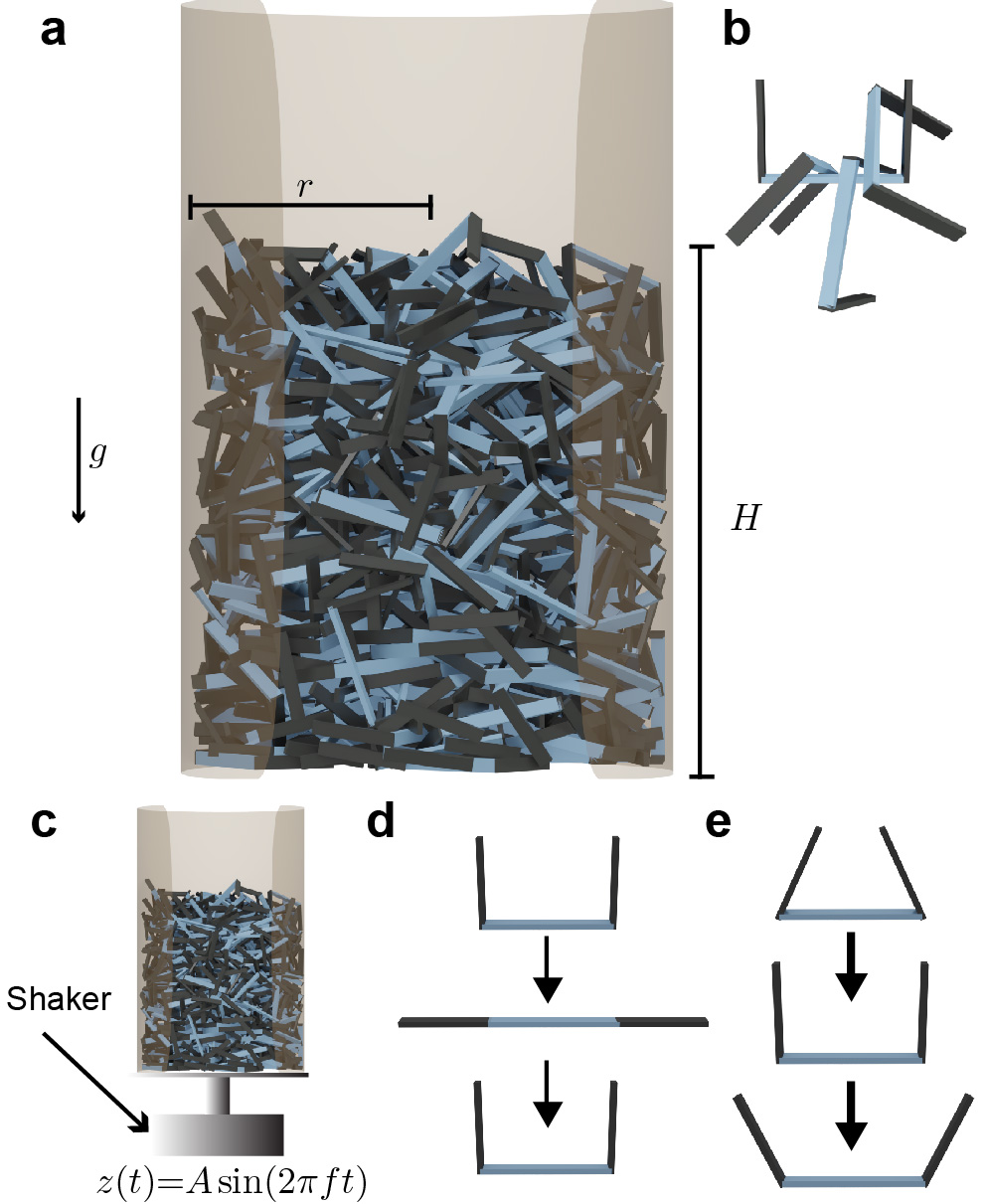}
    \caption[System entanglement procedures]{\textbf{Various smarticle activation procedures, internal and external.} (a) Render of simulated smarticle system deposited into a closed container of radius $r=2w$, one side is shown open to enhance visualization. (b) Rendering of 4 smarticles entangled together, here $\langle N\rangle=1.5$. (c) Externally oscillated procedure, particles are deposited into system and container is shaken sinusoidally, parallel to gravity, such that $z(t)=A \sin{(2\pi ft)}$. (d) In the shape-change procedure, smarticle barbs travel $\pi/2$ outwards and return back to the original position, this position change happens a single time in this procedure. (e) Internal oscillation procedure, particle barbs oscillate an amplitude $\theta$ centered around $(\alpha_1,\alpha_2)=(\pi/2,\pi/2)$.}
\label{fig:2a}
\end{figure}

Each simulation consisted of three phases: a deposition phase Fig.~\ref{fig:2a}(a), an activity phase, and a final testing phase for certain simulations. A relaxation period was added between each activity phase (when energy is added to the system) and testing phase. During this period, no activity is added to the system. We chose a relaxation period of 0.5 seconds, which was we determined to provide sufficient amount of time for the system to settle. In the deposition phase, smarticles were given a random initial and rotation position, and were released to fall by gravity into a hollow cylinder of radius $r=2w$. All barbs were held static at the position: $(\alpha_1,\alpha_2)=(\pi/2,\pi/2)$ during this phase (see Fig.~\ref{fig:simAndExpPics}(a)). The activity phase began once all smarticles were deposited in the system. This phase was responsible for changing the material properties of the collective by affecting the average entanglement between smarticles Fig.~\ref{fig:2a}(a,b). We define entanglement as the interpenetration of concave particles.

In the activity phase, activity was added to the system in one of three ways, where each active procedure represented a categorically different type of motion: an external sinusoidal oscillation of the confining cylinder (``externally oscillated'') Fig.~\ref{fig:2a}(c), a single cycle of a large angular change of each particles' barbs (``shape-change'') Fig.~\ref{fig:2a}(d), and finally, many small amplitude oscillations of each particles' barbs (``internally oscillated'') Fig.~\ref{fig:2a}(e). In the externally oscillated procedure, the container was shaken with a peak acceleration $\Gamma=2$ (in units of gravitational acceleration $g$) at $f = 30$~Hz and oscillated for 20~s (600 cycles) Fig.~\ref{fig:2a}(c). The shape-change procedure represented large amplitude movements. In shape-change simulations, the barbs were given 1.5 s to actuate from $(\alpha_1,\alpha_2)=(90^{\circ},90^{\circ})$ to $(\alpha_1,\alpha_2)=(0^{\circ},0^{\circ})$, then another $1.5$ s to actuate back to their original position $(\alpha_1,\alpha_2)=(90^{\circ},90^{\circ})$ (Fig.~\ref{fig:2a}(d)). The internally oscillated procedure represents small amplitude vibrational motion. The arc length traveled by the barbs' tips, for a given degree amplitude $\theta$, was constant for all $l/w$ in the internally oscillated trials. Each barb's tip displaced a distance equivalent to the arc length traveled by an arm of length $l$ with $l/w=0.7$ an amount $\theta=\pm [5^{\circ},30^{\circ}]/2$ (Fig.~\ref{fig:2a}(e)) from the original $(90^\circ,90^\circ)$ position. This ensured that the arc length traced by the barb tip was equal, for all $l/w$, at a single oscillation amplitude $\theta$. 
The activity phase in the internally oscillated procedure lasts 5 s before any subsequent procedures start. These various methods were used to encourage higher entanglement than that achieved from only deposition. Following the activity phase was a testing phase. 

In the testing phase, we performed a casting test, which examined qualitative material properties of a collective after smarticle activations. In summary, after smarticles were deposited in a container of a certain shape and activated, the container walls were removed for testing.

In all the simulations, the range of aspect ratios tested in the smarticle system for all procedures was  $l/w \in [0.4,1.1]$. For the internally oscillated system, the oscillation amplitude was $\theta \in [5^\circ,30^\circ]$.  

\begin{table}
\small
\caption[Smarticle chain simulation parameters]{simulation parameters}
\label{tab:simPars}       
\begin{tabular*}{0.48\textwidth}{@{\extracolsep{\fill}}ll}
\hline
parameter & value  \\
\hline
$w$ & 1.17 cm  \\
$dt$ & $2 \times 10^{-4}$ s  \\
$f$ & 30 Hz \\
$\Gamma$ & 2 \\
$\rho$ & 7850 kg/m$^3$\\
$\tau$ & 1  \\
$\omega$ &  6 rads$/$s\\
$t_1$ & 1.27 mm\\
$t_2$ & 0.5 mm \\
$\mu_{particle-wall}$ & 0.4\\
$\mu_{particle-particle}$ & 0.4\\
$r$ & 2$w$\\
$H_T(t_i)$ & $20 t_2$\\
$H_B(t_i)$ & $10 t_2$\\
\hline
\end{tabular*}
\end{table}

\subsection{Robophysical experiments of smarticles}
To test the physical attributes of a configurable chain of non-convex granular materials, three-link robots were built~\cite{Savoie2018,Savoie2019}. These robots, called ``smarticles'', are shown in Fig.~\ref{fig:simAndExpPics}(c,d,f). The arms are controlled by two servos (Power HD, HD-1440A) to a precision of ($<1^{\circ})$ and with an accuracy of $\pm6^{\circ}$. All processing and servo control is handled by an Arduino Pro Mini 328 - 3.3 V 8 MHz. Each robot is powered by a $3.7$~V $150$~mAh $30$~C LiPo battery (Venom;Rathdrum, ID.).

Due to the size of the servos and the thickness of the body Fig.~\ref{fig:simAndExpPics}(e,f), the dimensions of the smarticles prevented it from performing 3D entangling tests mentioned in Sec.~\ref{sec:smartSim}; the thickness of a robotic smarticle's center link limits the number of simultaneous particle interpenetrations, Fig.~\ref{fig:simAndExpPics}(e) and Fig.~\ref{fig:2a}(b). 
Despite this shortcoming, we found interesting results for these smarticle robots in 2D tests which depend on their non-convex nature. 

The u-shaped particles, when strained, produce auxetic behavior because of their concave shape. Furthermore, the strength of a ``chain'' of smarticles, defined by its resistance to fracturing under strain, is affected by the confinement of the chain. 

\subsection{Comparing multibody simulations to nature}
\label{sec:worms}
We compared smarticle simulations to emergent physically-entangled behaviors demonstrated by blackworms. Outside of their natural granular or detritivorous habitat, these worms entangle together forming a highly dense worm blob (Fig.1d) due to their thigmotactic behavior~\cite{2015Timm, 1990Drewes}. Furthermore, depending on the oxygenation of their surrounding, blackworms can disentangle their tails from the collective and lift it up to supplement respiration in hypoxic, or low dissolved oxygen (DO), conditions~\cite{1990Drewes, 2015Timm}. By altering the dissolved oxygen in a container, we can vary the entanglement strength displayed by a worm blob and observe its emergent dynamics. We define low and high DO as <2 mg/L and >8 mg/L, respectively. 

\subsubsection{Animals}

We purchased, reared, and habituated blackworms as described in Tuazon, et al. The experimental setup and data analysis is also the same \cite{2022Tuazon}. For the toppling experiment, blackworms were placed in a 250 mL cell culture flask with 100 mL of filtered water. We used MATLAB to generate the kymographs that tracts its center-of-mass (CoM) height as a function of time. Using blackworms for experiments do not require approval by an institutional animal care committee.


\section{Results and Discussion}
\subsection{Simulation}
\label{sec:sim}
\subsubsection{Packing fraction for an externally forced smarticle collective}

\begin{figure}[!ht] 
\centering
    \includegraphics[width=\hsize]{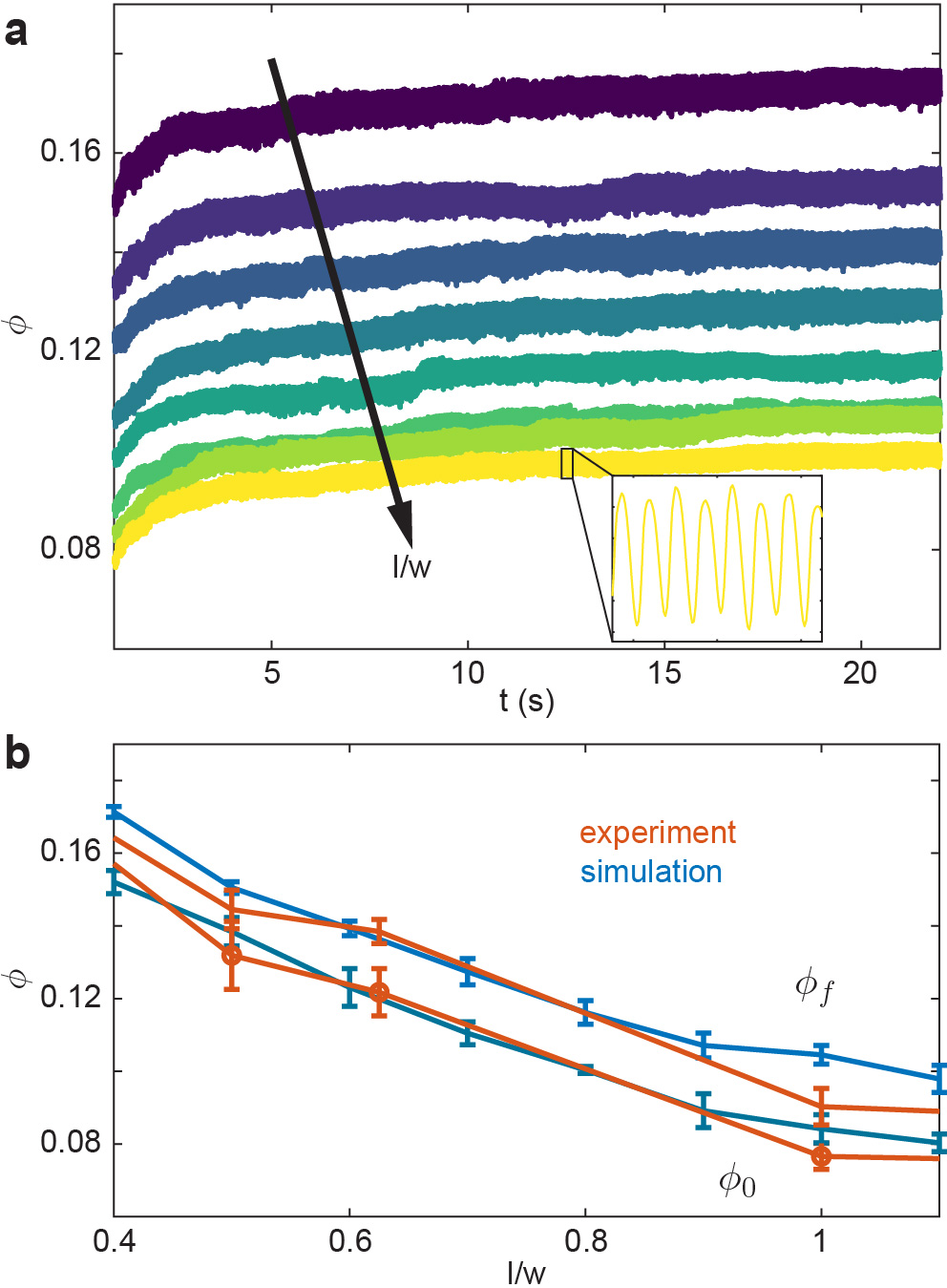}
    \caption[External oscillation packing fraction and simulation calibration]{\textbf{Packing fraction and packing evolution as a function of smarticle aspect ratio.} (a) $\phi$ vs. time for various smarticles aspect ratios. The inset is a zoomed in version of the data from $l/w=1.1$ over a 0.25 s domain. 
    (b) $\phi$ as a function of aspect ratio in experiment (orange) and simulation (blue). The bottom two curves represent the volume fraction before activity was added to the system $\phi_0$ in both simulation and experiment. The top two curves are the volume fraction after external forcing $\phi_f$. The orange curves were taken from data used in Gravish et al.~\cite{Gravish2012}}.
    \label{fig:externalPhi}
\end{figure}

We began by testing how smarticle collectives pack when subjected to an external vibration. In granular material studies, compaction, or volume fraction, is a commonly measured material property~\cite{Callister2014}. Volume fraction, or $\phi$, is defined by $\phi=V_p/V$, where $V_p$ is the solid particle volume and $V$ is the volume of the collective. $\phi$ can greatly affect how a granular collective reacts to external perturbations~\cite{Gravish2014,Agarwal2019}. For a non-ordered system of spherical particles, $\phi$ will generally be in the range between random loose packing (rlp) and random close packing (rcp) values. $\phi_{rlp}\simeq 0.56$ is the lowest density arrangement of spheres that is capable of enduring stresses~\cite{Scott1969,Andreotti2013}, and $\phi_{rcp}\simeq0.64$ is the highest possible value for randomly packed systems~\cite{Onoda1990,Andreotti2013}. The highest possible packing for ordered mono-disperse collections is  $\phi_{max}=\frac{\pi}{3\sqrt{2}}\simeq 0.74$~\cite{wells1991}. 

We tested how $\phi$ evolves for a collective undergoing external oscillation. Fig.~\ref{fig:externalPhi}(a) shows how $\phi$ evolved over time for different aspect ratios. As $l/w$ increases, the initial and final volume fraction, $\phi_0$ and $\phi_f$ respectively, also increases. The inset in Fig.~\ref{fig:externalPhi}(a) shows $l/w=1.1$ with $t\in [12.375 s,12.625 s]$, the oscillations in the inset are consistent with the applied oscillation frequency $f=30$ Hz. Furthermore, the rate at which $\phi$ increases, as a function of $l/w$, is faster than linear, which is illustrated more clearly in Fig.~\ref{fig:externalPhi}(b). This is true for both the $\phi_0$ and $\phi_f$. Fig.~\ref{fig:externalPhi}(b) shows a comparison of our simulation to results from a similar experimental system~\cite{Gravish2012}. Both $\phi_0$ and $\phi_f$ are monotonically decreasing functions of $l/w$. Our simulation results agree with the experimental system.

\begin{figure*}[ht]
\centering
\includegraphics[width= 174mm]{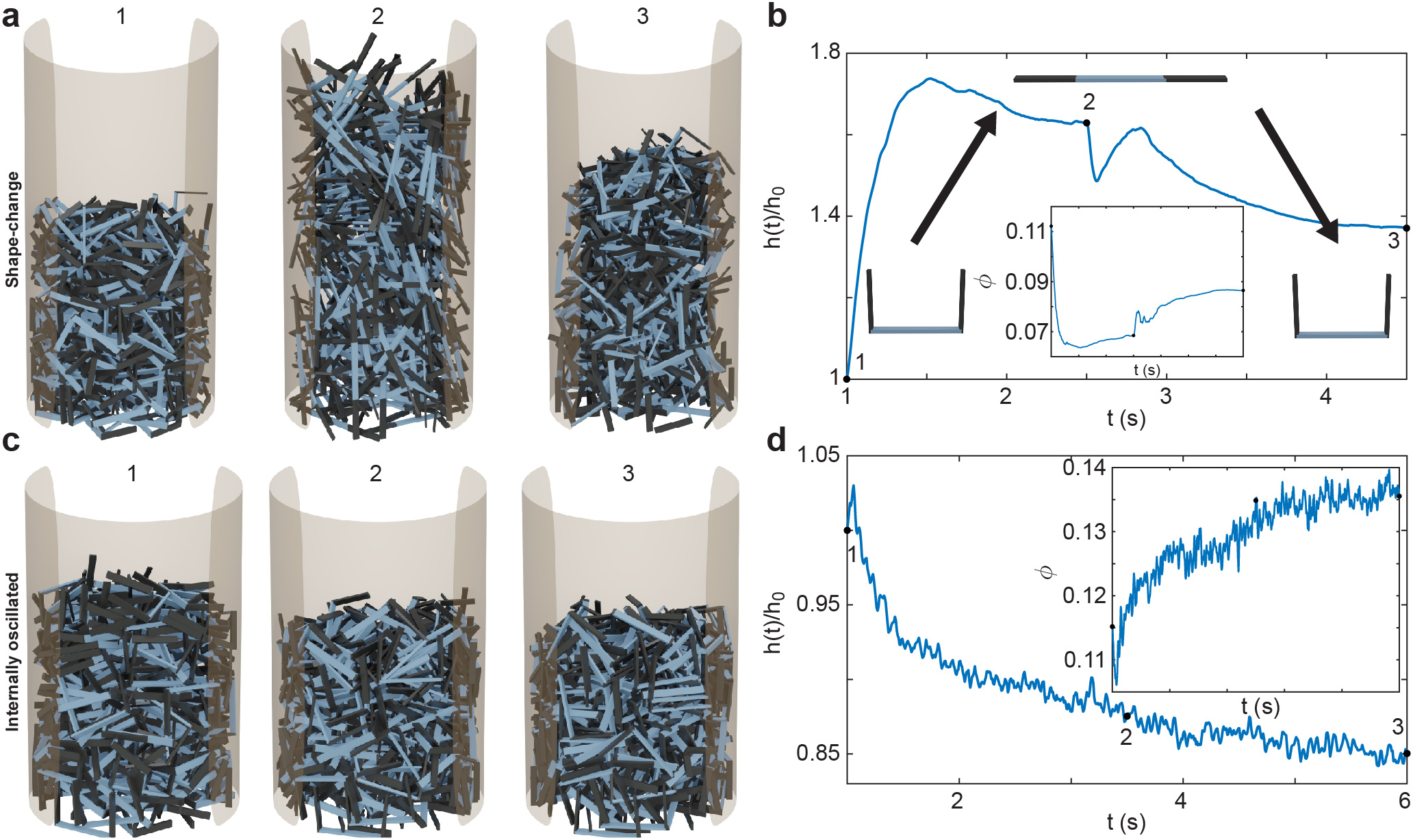}
\caption[Packing behavior for different procedures]{\textbf{Packing behavior distinctions between two different active procedures.} (a) Renders of a shape-change trial at three times. (b) The center-of-mass (CoM), $\frac{h(t)}{h_0}$,  in the z-plane of the shape-change collective, the times of the renders in (1,2,3) are highlighted with black points $[1s, 2.5s, 4.5s]$. (c) Renders of an internally oscillated trial at three times. (d) The CoM in the z-plane of the internally oscillated collective, the times of the renders in (1,2,3) are highlighted with black points $[1s, 3.5s, 6s]$.}
\label{fig:activePacking}
\end{figure*}

\subsubsection{Packing fraction for a smarticle collectives with internal degrees-of-freedom}

In the following trials, we examine packing evolution for internally activated smarticles. In Fig.~\ref{fig:activePacking} we introduce the effects of internal activations by showing the time evolution of a single trial for the different activation procedures. In Fig.~\ref{fig:activePacking}, we show three different frames from a shape-change activation procedure. In Fig.~\ref{fig:activePacking}(b), we show both the CoM height in addition to the inset containing the volume fraction. In the shape-change procedure, the final collective's shape fills the container differently with time as shown in Fig.~\ref{fig:activePacking}(a.1-3). The dynamics of the procedure are better captured by both the CoM height and $\phi$ evolution in the shape-change procedure. As the shape-change is initiated, we see a rapid increase of $h$. This increase results from the smarticles straightening. As the second shape-change phase happens, there is an initial decrease in height as smarticles curl into the u-shape. 

The smarticle collective's height does not decrease to the initial height of $h_0$, despite all constituents being in the same shape as deposited. Since the number of smarticles does not change during a trial, the change in height suggests that the procedure has changed the macroscopic state. After performing the shape-change procedure, the smarticles were not as well-molded or sculpted to the container as before. This suggests that the shape-change activation tends to force the smarticles away from the container wall as they transition back to the final u-shape. This inwardly-directed movement during the activation indicates an inwardly pointing attractive force. We suspect that the origin of the force arises from the geometry of the smarticle's final shape, as well as from the confining walls. The now excluded volume between the outer walls and curled smarticles drives the collective upwards, increasing its height after the initial drop after point 2 in Fig.~\ref{fig:activePacking}(b). 

The internally oscillated smarticles affect the collective differently than the shape-change procedure, Fig.~\ref{fig:activePacking}(c). As the smarticles vibrate their barbs, the smarticles interact with their neighbors, which causes them to rotate and displace. This allows particles to siphon into empty spaces that may exist between neighbors. Since gravitational force points downward, most rearrangement tends to lead to the compaction of particles until an equivalent $\phi_{rcp}$ for the smarticle shape is approached. As the general action fills voids, the collective tends to decrease in height and becomes denser. This increase in density means that the smarticle collective is molding more closely to the container boundary; thereby producing a final height of less than $h_0$. In Fig.~\ref{fig:activePacking}(d), we see that the majority of the compaction happens rapidly, in fact, the first 50\% of the final compaction is reached after only $\approx12\%$ of the total activation time has elapsed.

\begin{figure}[!t] 
\centering
    \includegraphics[width=\hsize]{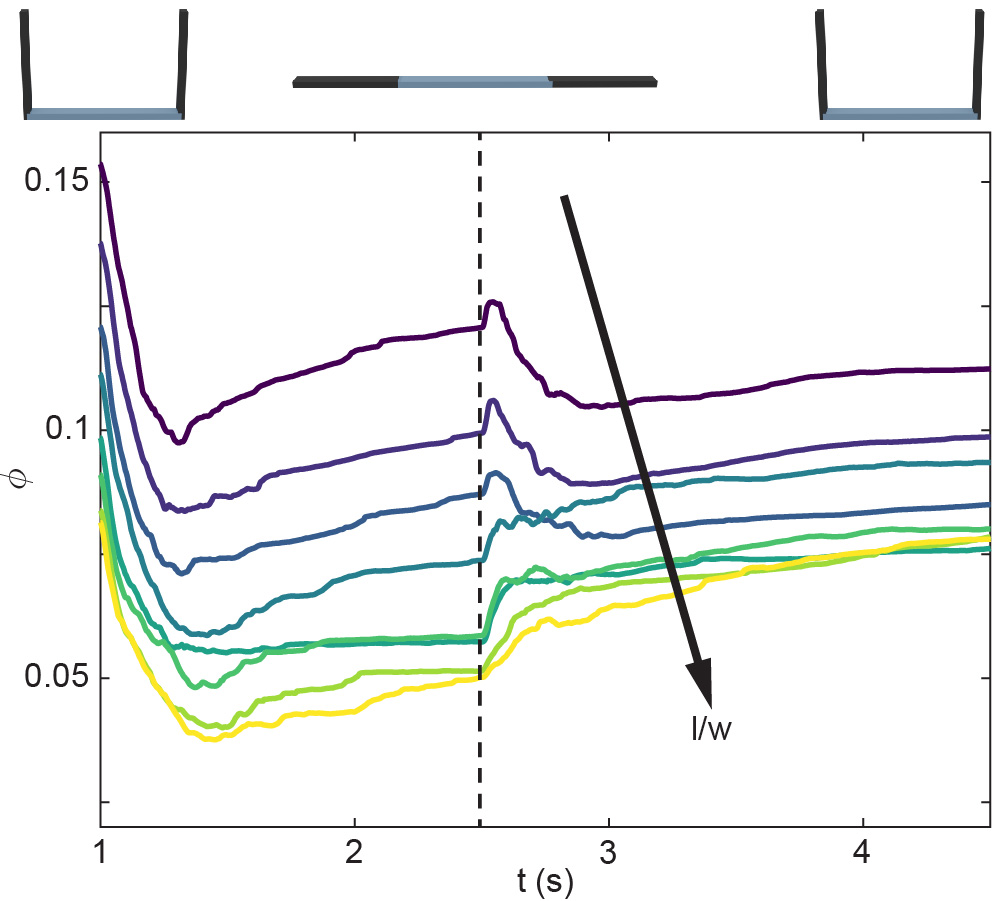}
    \caption[Evolution of packing fraction at different aspect ratios for the shape-change procedure]{\textbf{Evolution of $\phi$ at different $l/w$ for the shape-change procedure.} Time evolution of $\phi$ for varying $l/w$ performing the shape-change procedure. The vertical dashed line represents denotes when the “straight” to “u” configuration change begins.}
    \hfill \break
    \label{fig:SCphiEvolution}
\end{figure}

Next, we look at how the geometry of the smarticles, as well as the amplitude of their movements, affects the packing state of each system. In the shape-change procedure, $\phi$ generally decreases as the aspect ratio increases (Fig.~\ref{fig:SCphiEvolution}). Past $l/w>0.6$, $\phi$ monotonically increases (rather than increasing for a short time) then decreases. This behavior suggests at the existence of a critical packing fraction, which likely existing between $0.7<l/w_{crit}<0.6$.

\begin{figure}[t!] 
\centering
    \includegraphics[width=\hsize]{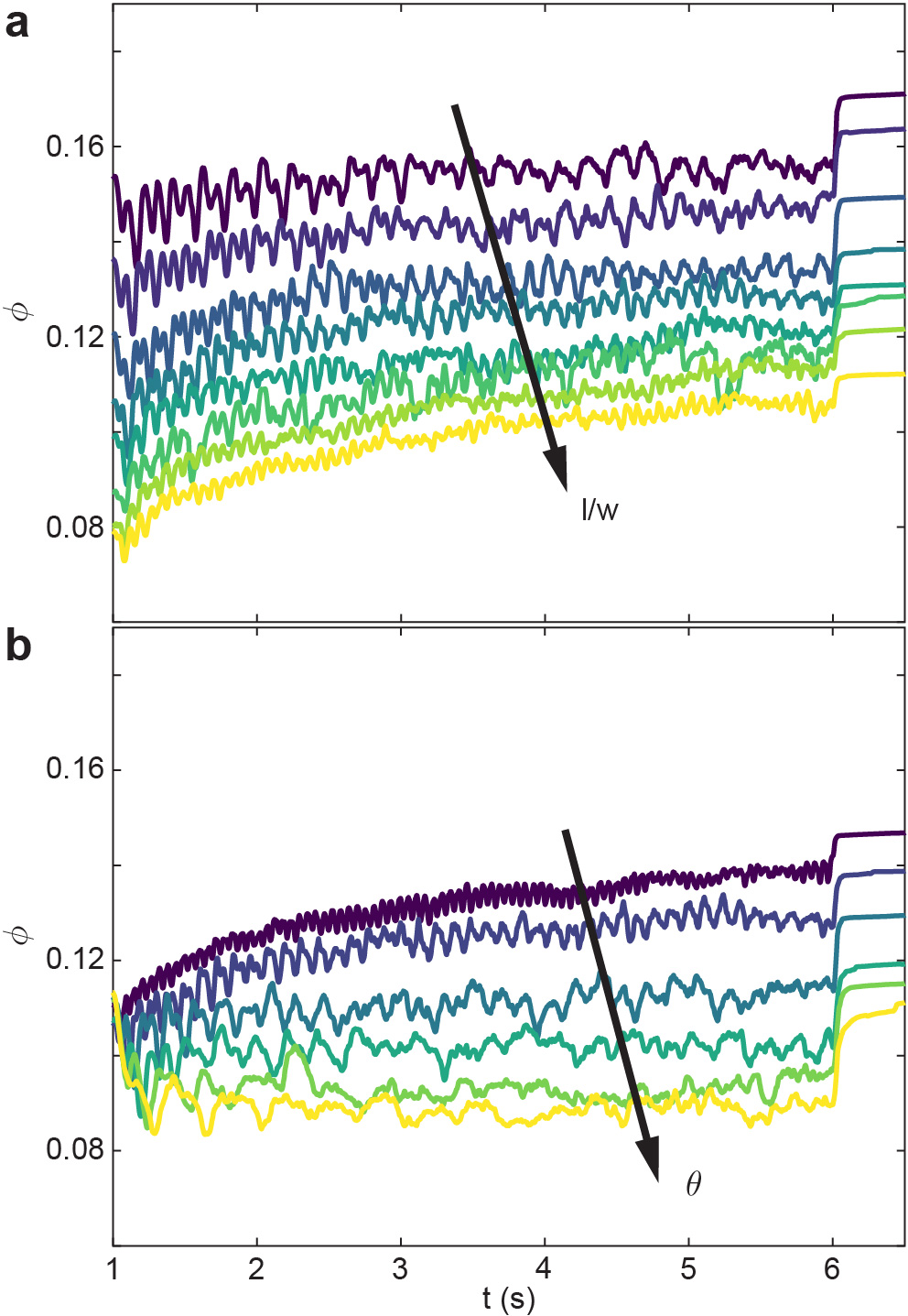}
    \caption[Evolution of packing fraction at different aspect ratios]{\textbf{Evolution of $\phi$ at different $l/w$ for the internally oscillated system.} Time evolution of $\phi$ for for various $l/w$ (a) and $\theta$ (b) for the internally oscillated procedure. (a) $l/w$ is varied  while $\theta=10^\circ$ is held constant. (b) $\theta$ is varied while $l/w=0.7$ is held constant.}
    \label{fig:VibPhiEvolution}
\end{figure}

The internally oscillated procedure is affected similarly as the shape-change with regards to the aspect ratio $l/w$. In Fig.~\ref{fig:VibPhiEvolution}(a) $\theta=10^\circ$ is held constant, and just as with the shape-change procedure, $\phi$ of the internally oscillated procedure decreases with increasing $l/w$. $\Delta \phi=\phi_f-\phi_0$ also increases with increasing $l/w$. For a given smarticle in a collective, as the barbs get longer, the number of neighboring smarticles it interacts with will generally increase. During internally-activated procedures, longer barbs tend to create more overall motion in, and of, the collective. The increased motion, or collective ``temperature'', tends to rearrange smarticles when space is available. For the simulation, at a given $\theta$, the arc length traveled by barb tips is constant for all $l/w$. This explains why the oscillations in $\phi$ tend to decrease as $l/w$ the oscillation frequency increases as $l/w$ increases.

Given a constant $l/w=0.7$, in Fig.~\ref{fig:VibPhiEvolution}(b) we vary oscillation amplitude $\theta$. We find that $\phi$ decreases as $\theta$ increases. As oscillation amplitude increases, larger vacancies form between neighbors, but the larger $\theta$ hinders flow into the temporary vacancies made by the barb motion. While $\theta$ increases, $\Delta \phi$ decreases as well.

\begin{figure}[t!] 
\centering
    \includegraphics[width=\hsize]{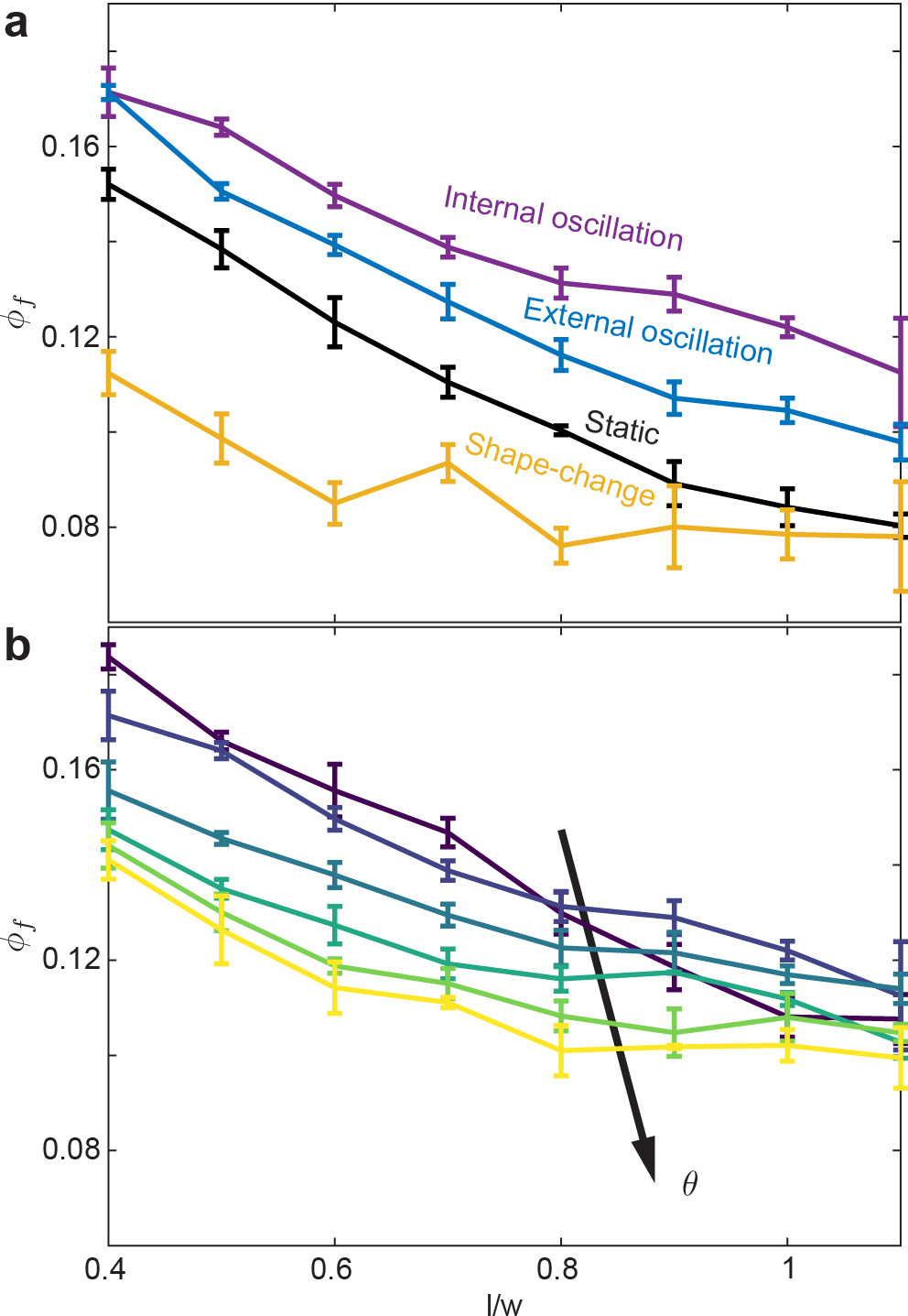}
    \caption[final packing fraction for all system procedures]{\textbf{$\phi_f$ fraction for the various procedures.} (a) $\phi_f$ versus $l/w$ for all procedures. For the internally oscillated procedure, $\theta=10^\circ$ was used. (b) $\phi_f$ versus aspect ratio for the internally oscillated system, each line is has a different $\theta$.}
    \label{fig:phiCompareAll}
\end{figure}

Finally we compare how $\phi_f$ varies with $l/w$ for all activation procedures (Fig.~\ref{fig:phiCompareAll}(a-b)). The internally oscillated procedure tends to pack more densely, whereas shape-change tends to pack the least densely Fig.~\ref{fig:phiCompareAll}(a). Among various oscillation amplitudes, lower amplitudes tend to pack more tightly than larger amplitudes Fig.~\ref{fig:phiCompareAll}(b). Indeed, the largest amplitude shows only a marginal increase in $\phi$ compared to the static system, where there is no smarticle activation after the deposition phase (Fig.~\ref{fig:phiCompareAll}(a)).

\subsubsection{Energy comparison between procedures}
To clarify the cost of each procedure in terms of energy input to packing output, we measured the energy necessary to complete each procedure. This test was performed for smarticles of varying aspect ratio. Each simulation was performed three times for each preparation style at each aspect ratio and oscillation amplitude when applicable.
To calculate the energy of the active procedures,  the following calculation was performed at each time step,
\begin{equation} 
E=\int_{\alpha_1(t)}^{\alpha_1(t+dt)}\tau_1 d\alpha_1 + \int_{\alpha_2(t)}^{\alpha_2(t+dt)}\tau_2 d\alpha_2
\end{equation}
where the subscripts correspond to the barbs on each smarticle. Here, $\tau_i$ is the motor reaction torque on barb $i$. Since the energy is added into the system for the externally oscillated system,  we sum

\begin{equation} 
E=\int_{z(t)}^{z(t+dt)}F(t)dz
\end{equation}
over each time step during the shaking process. Here $F(t)$ is the reaction force on the linear actuator (which controls the height of the container with the smarticles) motor at time $t$, and $z(t)$ is the distance traveled by the cylinder at time $t$. The contribution of the container's weight was removed from the force calculation.

\begin{figure}[!ht] 
\centering
    \includegraphics[width=\hsize]{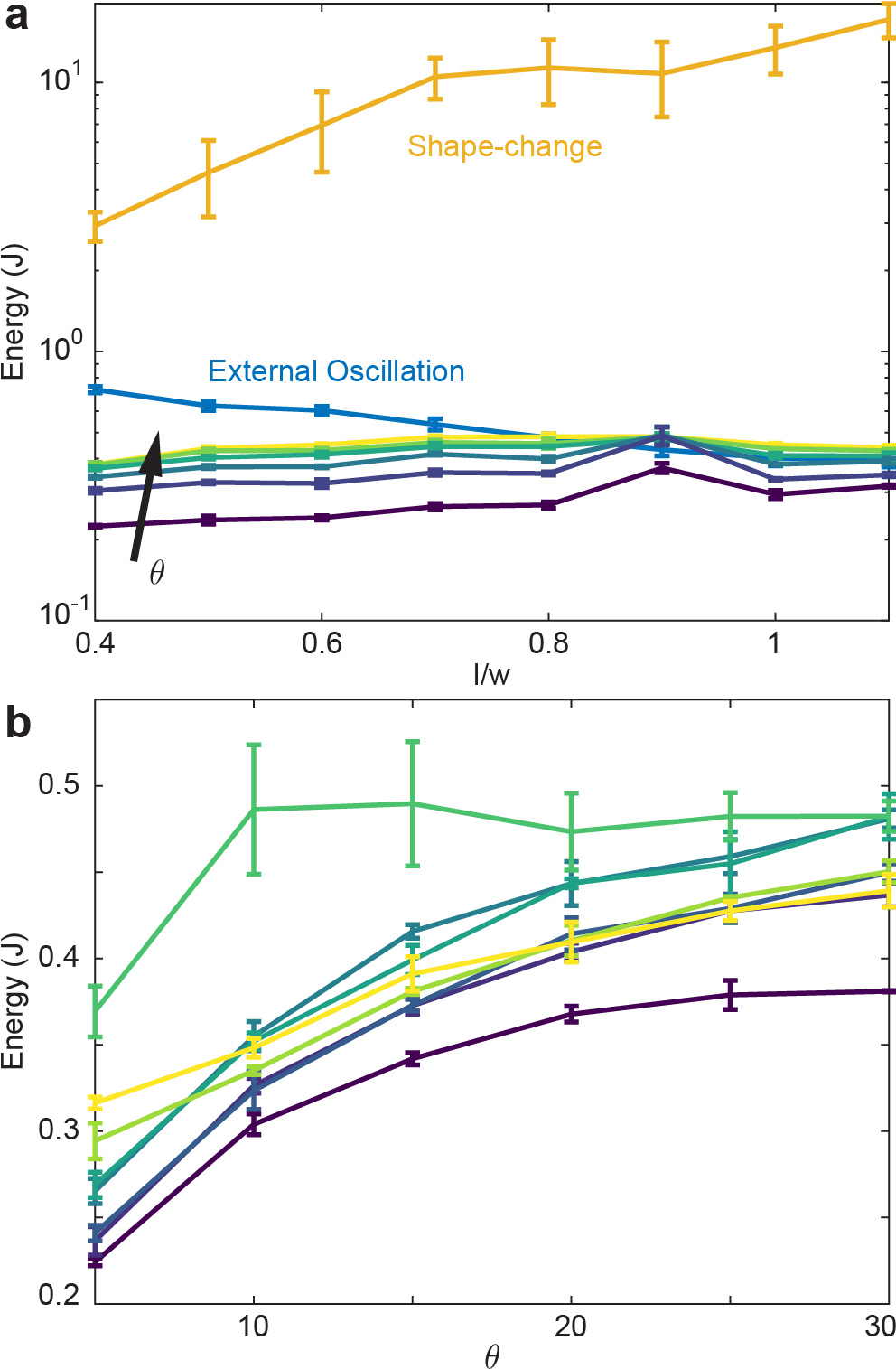} 
    \caption[Comparing energy usage between procedures]{\textbf{Comparing energy usage between procedures.} (a) Comparisons of energy output for different procedures as a function aspect ratio. The internally oscillated data are the unlabeled lines where $\theta$ is varied. All data points are averaged from three trials. (b) Energy usage for the internal oscillation procedure as a function of $\theta$ with a constant $l/w=0.7$.
    }
    \label{fig:sysCompare}
\end{figure}

Fig.~\ref{fig:sysCompare}(a) reveals that the shape-change procedure requires the most energy to perform, regardless of the aspect ratio. Indeed, shape-change required greater than an order of magnitude increase in energy compared to the internally oscillated procedure, which required generally the lowest amount of energy. Plotted in Fig.~\ref{fig:sysCompare}(a) are all oscillation amplitudes in the internally oscillated procedure. As $\theta$ increases, the energy required to perform the procedure also increases. Similarly, the energy to perform the procedure increases with $l/w$ with the exception of local maxima at $l/w=0.9$. 
Plotted in Fig.~\ref{fig:sysCompare}(b) is the energy as a function of oscillation amplitude for various aspect ratios. Apart from
$l/w=0.9$, energy tends to increase with amplitude. Next, given each procedure's energy expenditure, we measure how the average entanglement in the collective varies for each procedure. 

\begin{figure}[!] 
\centering
    \includegraphics[width=\hsize]{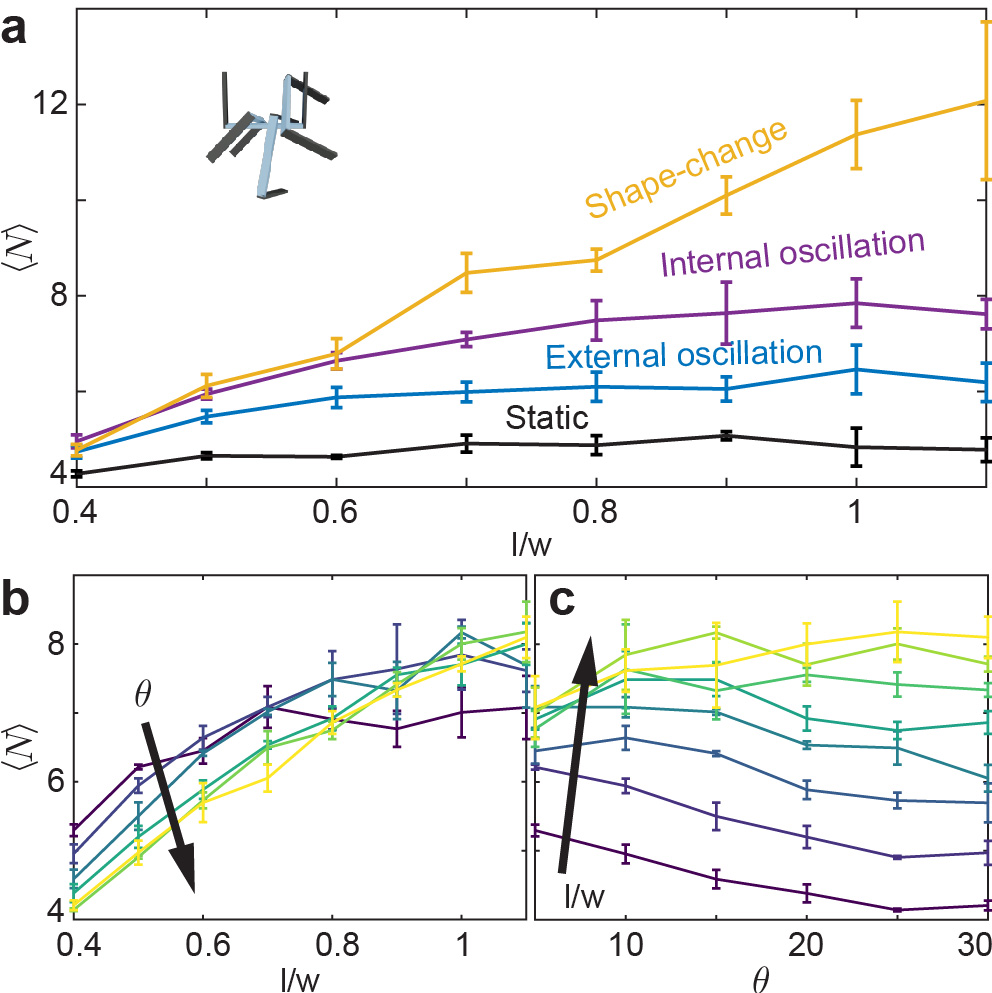}
    \caption[Comparing entangling procedures]{\textbf{Comparing entangling procedures.} (a) Comparisons of entanglements for all procedures at various aspect ratios. The internal oscillation line for each point has an amplitude $\theta=10^{\circ}$. (b) Entanglement with respect to $l/w$ for the internally oscillated system. (c) Entanglement as a function of $\theta$ for internally oscillated systems.}
\label{fig:systemEntangles}
\end{figure}

\subsubsection{Entanglement}
Since entanglement is affected by aspect ratio in externally activated procedures, we examine how it changes for internally activated procedures. After the collective has performed a procedure, the average number of entanglements, $\langle N\rangle$, between smarticles was measured. $\langle N\rangle$ was measured for the non-actuated (or static) procedure, the shape-change procedure, and the internally and externally oscillated procedures. $\langle N\rangle$ was determined as follows: for each smarticle A, there exists a plane defined by its 3 links. The number of smarticle links, not belonging to A, which intercepts smarticle A's plane is $N$. The mean of all $N$, over all smarticles in a single timestep, is $\langle N \rangle$ (see Fig.~\ref{fig:systemEntangles}(a) inset).

While entanglement is largely independent of aspect ratio in the static, or inactive, procedure, it is more dependent on aspect ratio in the active procedures Fig.~\ref{fig:systemEntangles}(a). Apart from the shape-change procedure, all other active procedures tend to display a maximum of $l/w \approx 1.0$. As $l/w$ increases, the barbs increase in size, thereby requiring more movement between two entangled components to become disentangled. Larger barbs make disentangling more difficult, this is true for an even greater degree in confined systems. Although smarticles with larger $l/w$ will remain entangled more readily, they are less likely to become entangled~\cite{Gravish2012}. The shape-change system does not have this issue: the entangling action happens during the straight to u-shape transition. During this time, a barb's length does not act to restrict a new entanglement from arising, as the smarticle shape starts from a convex (straight) shape. Therefore, higher values of $l/w$ increase the likelihood of entanglements happening in the shape-change system, as the area defining interpenetration is larger. 

\begin{figure*}[ht]
\centering
    \includegraphics[width= 174mm]{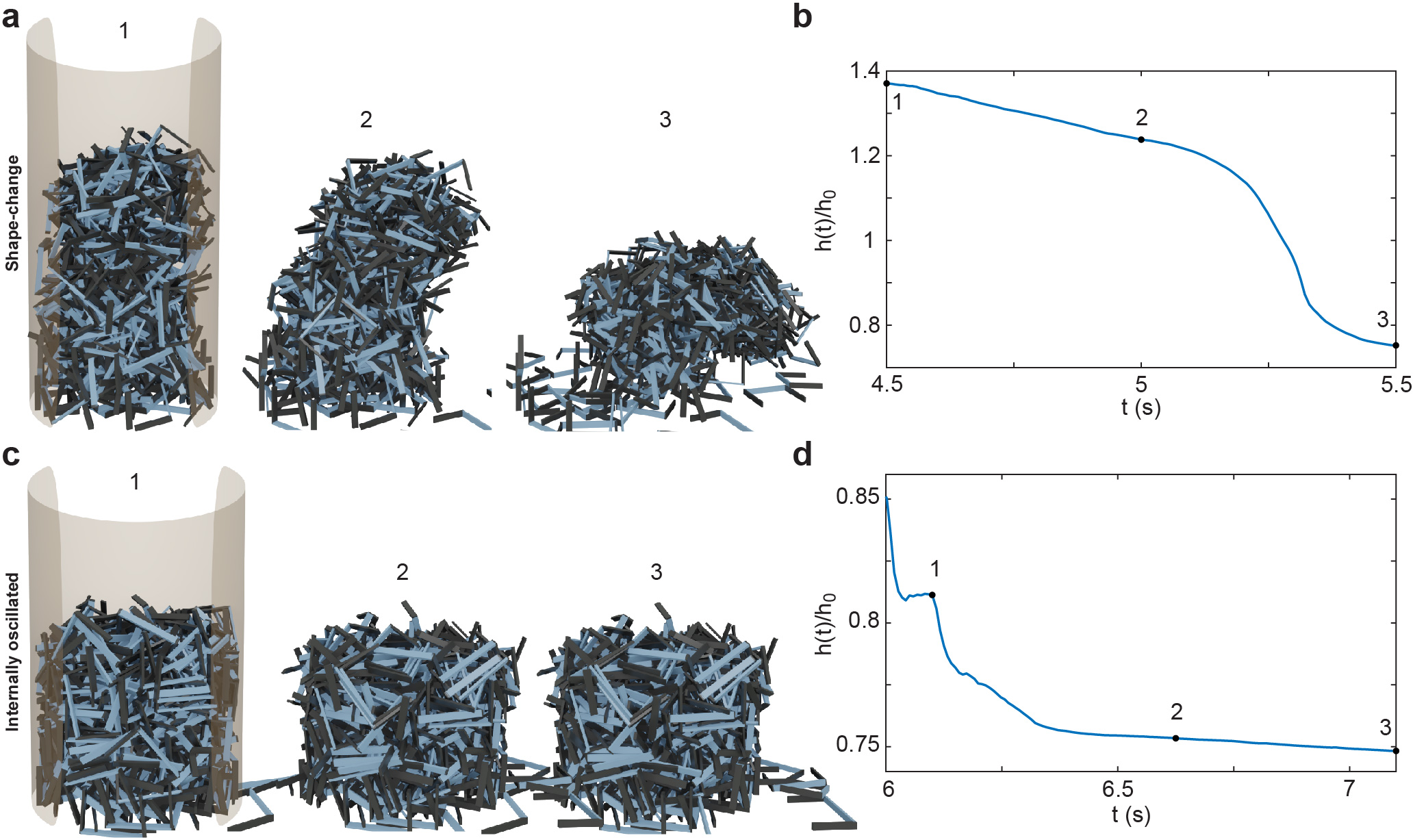}
    \caption[Falling behavior for different ]{\textbf{Melting behavior for active procedures.}(a,c) Renders of a shape-change trial and internally oscillated trial, respectively, at three instants. (b,d) The center of mass (CoM), in the z-plane, of the collective; the three renders from (a,c) are indicated with black points.}
    \label{fig:activeWallRemove}
\end{figure*}

\begin{figure*}[!hb] 
\centering
    \includegraphics[width= 174mm]{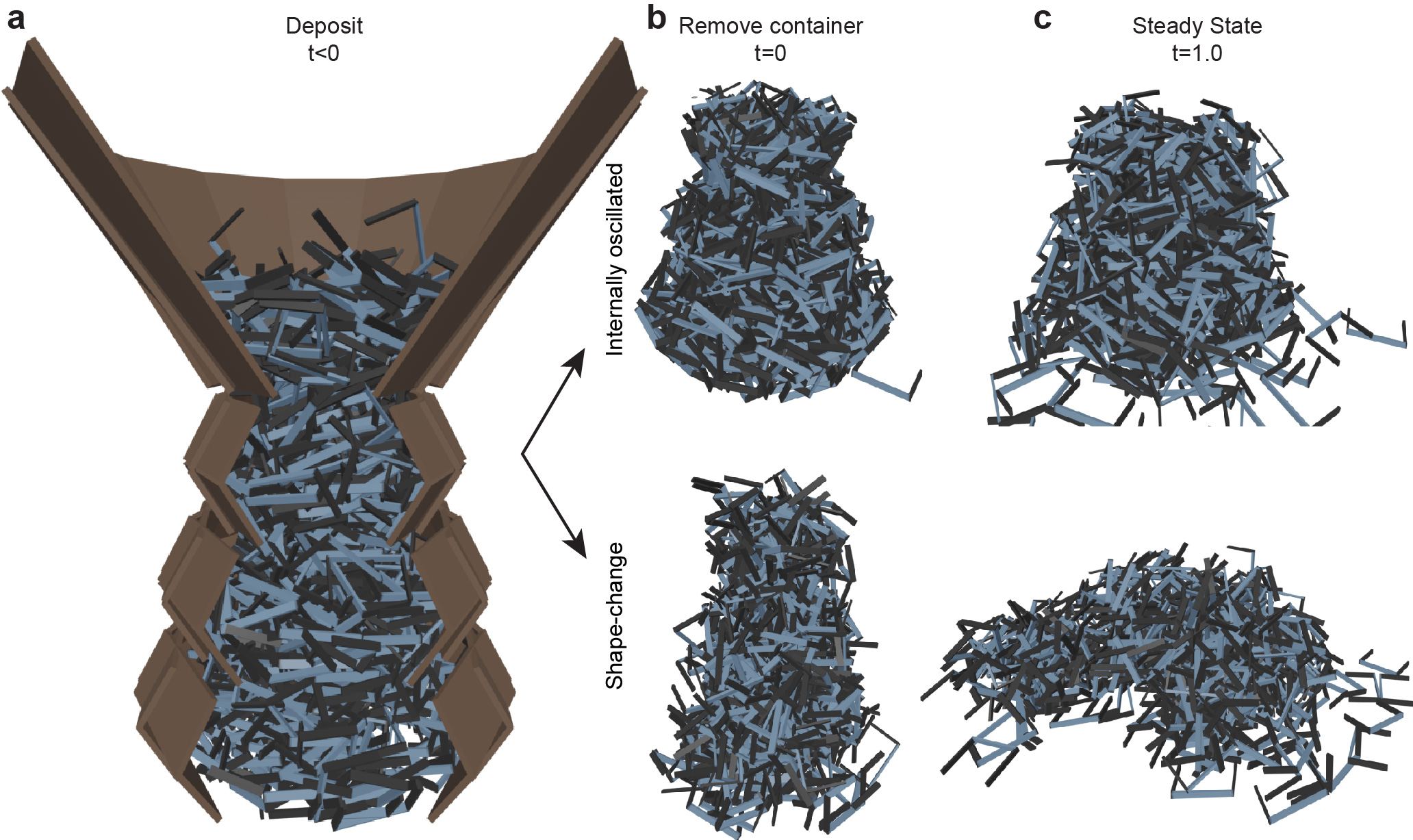}
    \caption[Smarticle aggregate sculpting capabilities]{\textbf{Smarticle aggregate sculpting capabilities.} (a) Particles are deposited a three-tiered system with $l/w=0.8$ with each tier smaller than the one below. In (b), after the activation state has occurred, (top is internally oscillated at $\theta=10^\circ$ and bottom is the shape-change) the outer walls are removed. After the particles are activated, the walls are removed. (c) The piles one second after the container is removed for the internally oscillated (top) and shape-change (bottom).
    }
    \label{fig:sculpting}
\end{figure*}

Now we examine how $\theta$ and $l/w$ affect $\langle N \rangle$ in the internally oscillated system. In these trials, we measured entanglements over $\theta=[5^\circ- 30^\circ]$ for all $l/w$ in Fig.~\ref{fig:systemEntangles}(b). At $l/w<0.7$, as $\theta$ increases, the number of entanglements decreases. At larger $l/w$, $\langle N\rangle$ becomes more independent of $l/w$. The functional form of entanglement, as a function of amplitude, changes as aspect ratio increase. At low aspect ratios, $l/w=[0.4-0.5]$, the curve is monotonically decreasing. However as $l/w$ increases, the curve is no longer monotonic: the peak forms between $\theta=[10^\circ-15^\circ$]. As barbs increase in size, the location of the peak increases too. Furthermore, the peak's relative amplitude to the rest of the curve also increases. We posit that this is related to the argument stated before: at low $l/w$, there is an increased likelihood to both entangle and to disentangle; the barbs can break the entanglements faster than new ones are generated. However, if the arms are longer, the same arc length traveled will not tend to disentangle an existing connection. The amplitude, as well as the energy threshold that is necessary to break pre-existing bonds, is larger. 

%

The static system (not shown in the energy plot) has zero energy input, and the $\langle N\rangle$ is the lowest for all aspect ratios measured. The internally and externally oscillated procedures have similar energy inputs; although the internally driven system is slightly lower. In the internally driven system, however, the number of entanglements measured after the deposition phase is noticeably higher. Aside from requiring different amounts of energy to prepare each procedure, each procedure also tends to impart different collective attributes. Since entangling allows a collective to support tensile loads, next we test how well each procedure allows a collective to both sculpt in a cast, and keep its shape after the cast is removed. 


\subsubsection{Melting behavior}
We check the qualitative differences between collections of smarticles after each procedure was performed and after the confining walls were removed. Inspired by heat-treatment annealing procedures in metallurgy and their effects on a metal's final hardness properties during fast and slow cooling, respectively, we test how shape-change and internal oscillation change the properties of the collective~\cite{Callister2014}. Furthermore, we study how these preparation procedures allowed cohesive smarticle collectives to be sculpted by the container.

After the deposition and activation of smarticle collectives, in Fig.~\ref{fig:activeWallRemove}(a-d) we removed the outer boundary and examine how the unconfined collective settles. The shape-change system, Fig.~\ref{fig:activeWallRemove}(a-b), does not break apart nor do the particles on the periphery disengage from the collective: the collective holds firmly together. Since shape-change narrows the base and increases the height, this procedure tends to produce less stable structures. Indeed, in the illustrated example, the structure falls onto its side in Fig.~\ref{fig:activeWallRemove}(a.3). The internally oscillated system reacts differently than the shape-change system when the confinement is removed. Since the internally oscillated procedure increases in $\phi$, lowering the height of the collective and keeping the initial width, the stability of the final collective is improved compared to the initial state after deposition. However, because the amount of entanglement was lower than in the shape-change procedure after the walls were removed, some smarticles on the outer edge slough off as they are not entangled tightly with neighbors in the structure. In the next section, we will look in more detail at structure forming using these two procedures.

\afterpage{\FloatBarrier}
\afterpage{\clearpage}

\subsubsection{Casting}
Here, we show how certain shapes are formed or sculpted through the activity of the smarticles in the collective. Starting in the u-shape position, we deposited smarticles into a three-tiered empty structure with each tier smaller than the one below (Fig.~\ref{fig:sculpting}(a). After depositing the smarticles, the smarticles are activated using the two aforementioned internal activation procedures: internal oscillation in Fig.~\ref{fig:sculpting}(b-top) and shape-change in Fig.~\ref{fig:sculpting}(b-lower). The container serves to define the final shape.

While activation procedures tend to compact a collection, they do so in distinct ways; these distinctions lead to the macroscopic differences visible in the remaining structure after the container is removed. Internal oscillation imparts the fluid-like property of being able to fill its container, and the activity tends to help smarticles flow into voids. While the container is filled more completely (as shown by the increase in $\phi$), this does not imply increases in entanglement. This is evident in Fig.~\ref{fig:sculpting}(b) in the instant the walls are removed. The system has the shape of the container but, shortly afterward, the overhangs tend to fall apart. For the shape-change procedure, while the particles become more compact, it is not due to an improved container space usage. The walls serve to give a basic tubular shape, but the particle structure consolidates inward and grows upward rather than outward. Previously, we showed the resulting effects caused by the number of entanglements produced by the different procedures (Fig.~\ref{fig:systemEntangles}). Shape-change produces a higher entanglement allowing particles to stay together more readily, but it does not sculpt to the container as easily as the internally oscillated procedure. After the activation procedure completion in the shape-change procedure, the system exhibits properties of a hardened structure, whereas the internally oscillated procedure tends to flow apart and is comparably softer, eroding at its surface after the removal of the walls.  We find analogies between the shape-change and internal oscillation procedure and rapid and slow quenching in metallic systems. In metals, rapid temperature reduction in the quenching process tends to leave a metal harder but more brittle, with a great deal of internal stress. Conversely, a slowly quenched metal will see a reduction in the internal stress as well as becoming softer~\cite{Callister2014}. These effects are further supported through quantitative tests in the simulated fracture tests performed in the next section.

\subsubsection{Tensile loading and fracture}

\begin{figure}[h!] 
\centering
\includegraphics[width=\hsize]{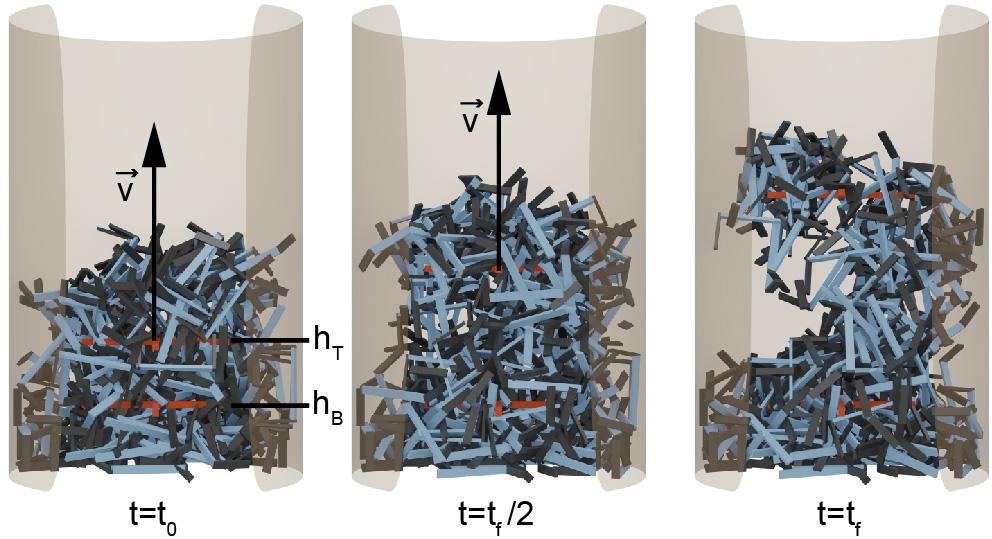}
\caption[Tensile loading in simulation]{\textbf{Tensile loading in simulation.} Measuring force as a function of entanglement procedure. Two hooks, colored red, are embedded in an already entangled pile. The force necessary to raise the top hook out of the pile is measured, while the bottom hook is kept fixed}
\label{fig:3a}
\end{figure}

To determine the relationship between entanglement and tensile loading ability, we measured the force required to raise a hook embedded in the various smarticle preparations as shown in Fig.~\ref{fig:3a}(a-c). Specifically, we measure the force as a function of aspect ratio, $F(l/w)$. The force required to raise the top hook was measured three times for each aspect ratio.

\begin{figure}[h!] 
\centering
\includegraphics[width=\hsize]{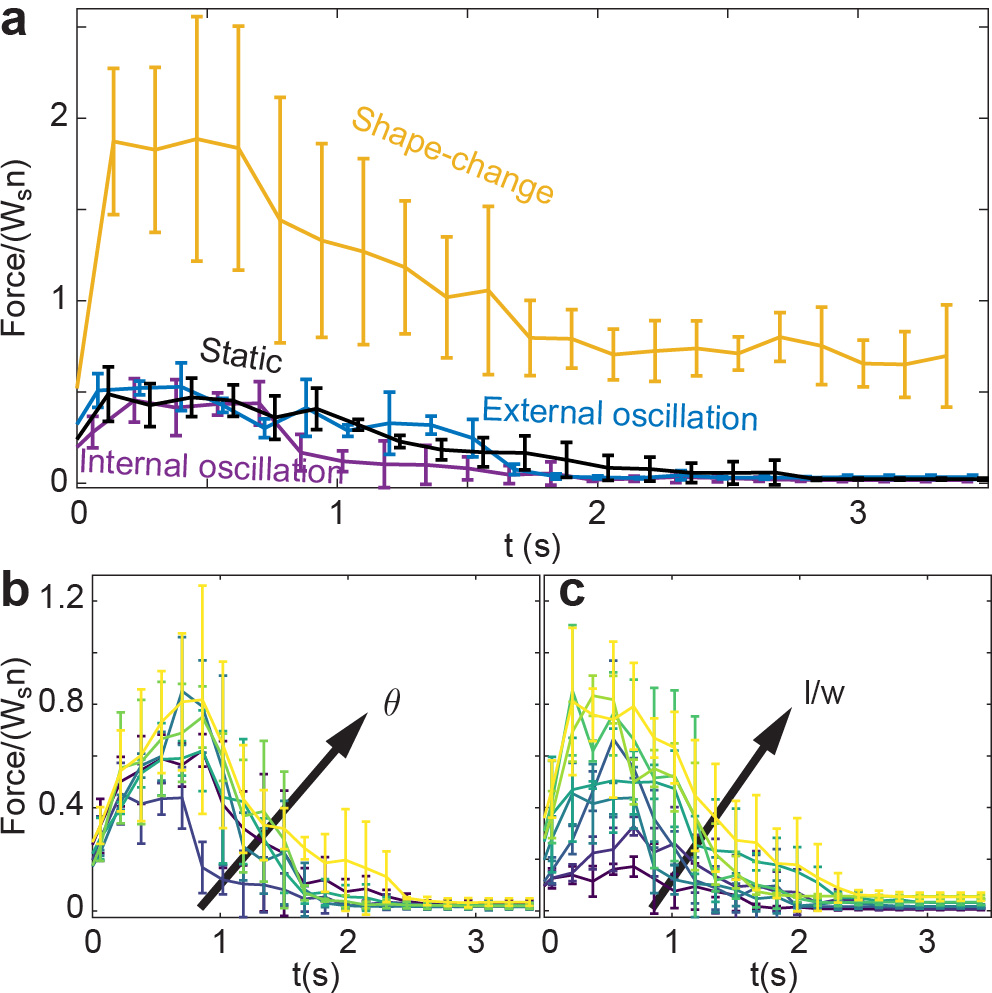}
\caption[Fracture force versus time for different systems]{\textbf{Tensile force measurements for the various preparation procedures.} Force shown here is a unitless quantity scaled by $W_s$, the weight of the smarticles, and $n$, the number of smarticles in the trial. Each line is averaged over 3 trials. (a) Force versus time for the various procedures at $l/w=0.7$, (b-c) Force versus time for the internally oscillated system.  In (b), $l/w=0.7$ and $\theta$ is varied, and (c) $\theta=10^\circ$ as $l/w$ varies.
}
\label{fig:forceDiffSys}
\end{figure}

To investigate how materials properties are affected, we performed vaious fracture tests. In Fig.~\ref{fig:forceDiffSys}(a) for the three procedures, the peak force output and the entanglements (Fig.~\ref{fig:systemEntangles}(a)) do not keep the same ordering. The shape-change system shows the highest peak force output, whereas internal oscillation displays the least amount of tensile strength. There is agreement between force generated as a function of internal oscillation as shown in Fig.~\ref{fig:forceDiffSys}(b) and in the number of entanglements for a given oscillation amplitude as shown in Fig.~\ref{fig:systemEntangles}(b). Using this simulated system, we found a connection between tensile strength, measured through a fracture test, and the number of entanglements. Next, we examine emergent properties arising in a planar fracture setup with smarticles in an experiment.

\subsection{Robophysical Experiment}
\begin{figure}[!h]
\centering
\includegraphics[width=\hsize]{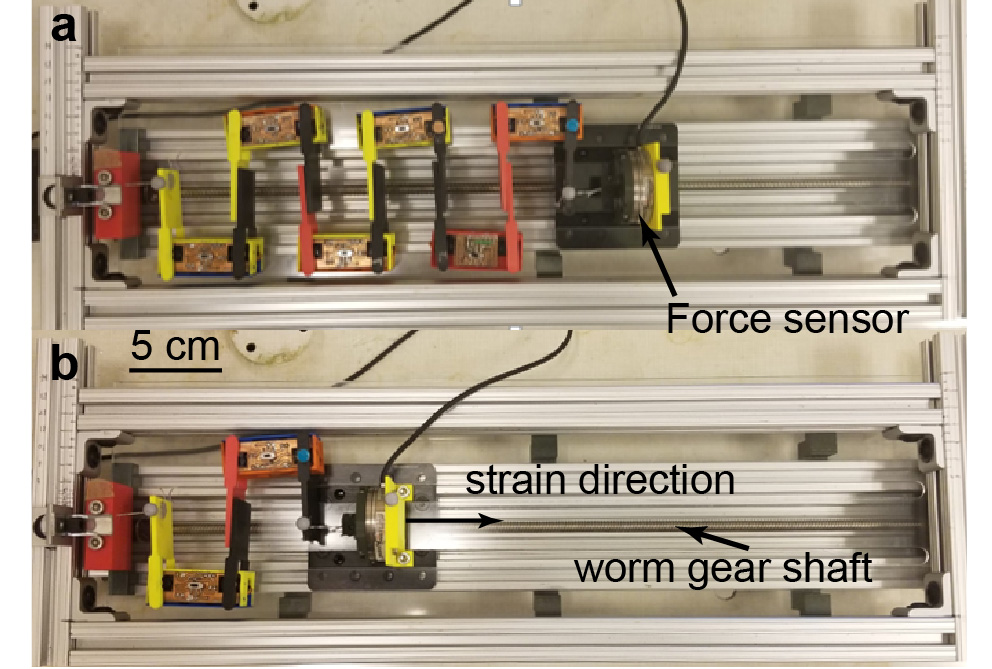}
\caption[n=6 and n=2 image]{\textbf{smarticle chains of varying size, $n=6$ and $n=2$.} 
(a) Top view of an $n=6$ system, (b) top view of an $n=2$ system.}
\label{fig:frac2-6}
\end{figure}

\begin{figure}[!h]
\centering
    \includegraphics[width=\hsize]{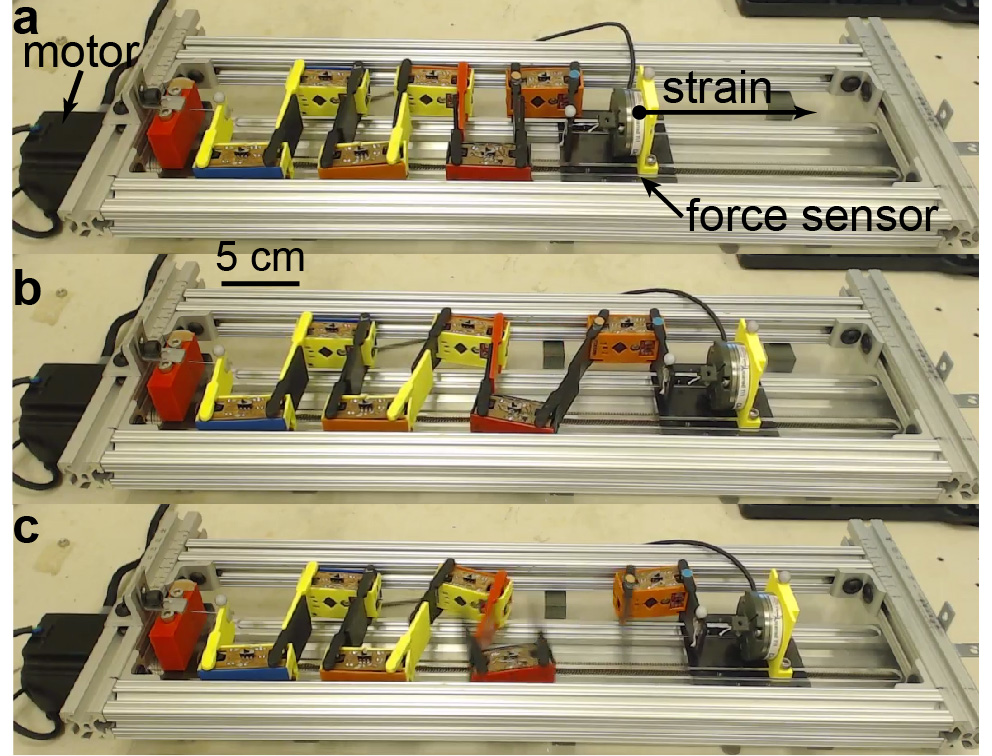}
    \caption[Fracture time sequence in experiment]{\textbf{Time sequence of chain fractures in experiment.} 
    (a-c) 3 snapshots of an $n=6$ chain of smarticle. (a) shows the chain before strain starts, (b) shows right before fracture, and (c) shows immediately after fracture.}
    \label{fig:timeSeq}
\end{figure}

Strain tests are a common method to test the elastic, plastic and yielding properties of materials. Here, we perform a strain test for a chain of smarticles. In all trials, smarticles were initially centered between the confining walls Fig.~\ref{fig:frac2-6}(a). The positions of each smarticle in the chain were randomized between each trial to account for any variance in the servos' strength due to manufacturing differences or general wear that may accumulate over time. Experiments are performed with two different amounts of smarticles, $n=[2,6]$. The chains were arranged as shown in Fig.~\ref{fig:frac2-6}(a,b) in a repeating $\sqcup-\sqcap-\sqcup-\sqcap$... pattern, where a ``$\sqcap$'' is the same shape as a ``$\sqcup$'' but rotated by $\pi$ rads. This pattern interlocks adjacent smarticles together. For chains of $n>1$, stress was transmitted between smarticles  via the entanglement between their barbs. The barbs on the ends of the chain that are not in contact with an adjacent smarticle were connected to the apparatus via a string. On one side the chain is a force sensor which is affixed to a static part of the apparatus' structure on the other side. The force sensor was connected on the side of the chain where the strain was imparted.  
The force sensor was a custom device with strain gauges in a full-Wheatstone bridge configuration. All force measurements, F(t), were sampled at 1 kHz. The strain was imparted by a modified linear actuator kit (OpenBuilds; Monroeville, NJ.) as shown in Fig.~\ref{fig:timeSeq}(a-c). All strains were performed at a strain-rate $v=[5$ mm/s$]$. The strain was measured by tracking infrared markers on the smarticles, where all positions and orientations of the smarticles were tracked using an infrared video recording hardware/software suite (OptiTrack; Corvallis, OR).

\begin{figure}
\centering
    \includegraphics[width=\hsize]{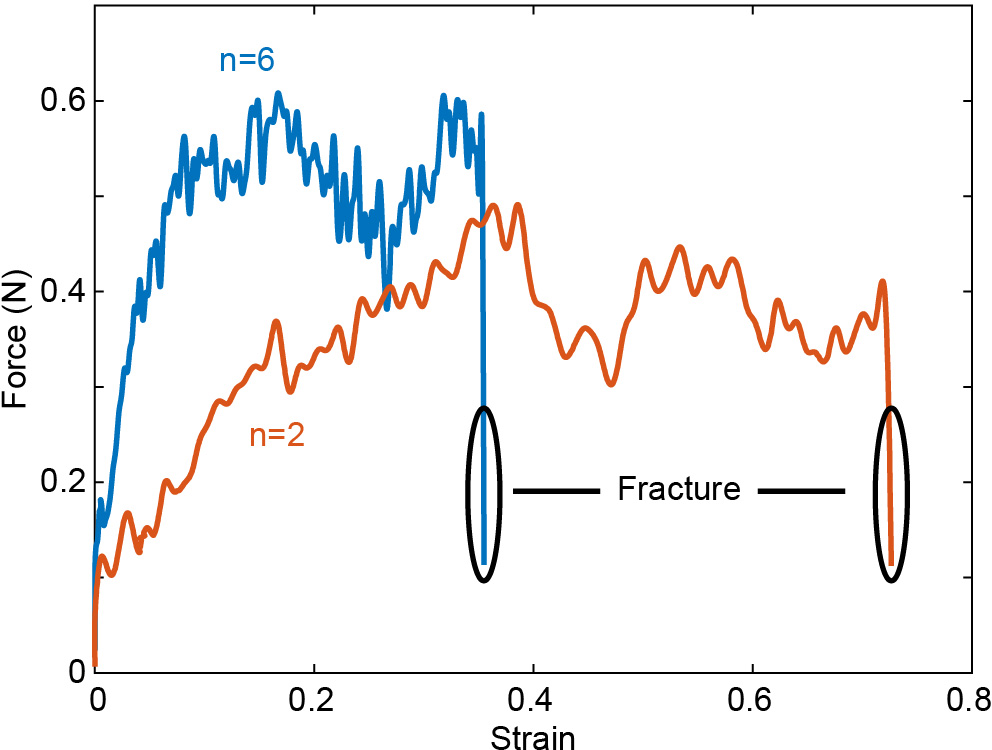}
    \caption[Single fracture trial for n=2 and n=6 smarticle chains]{\textbf{Single fracture trial for n=2 and n=6 smarticles.} Single trials of force as a function of strain for smarticle chain at different $n$, both strains were continued until the chain fractured.}
    \label{fig:fracture}
\end{figure}

\begin{figure}[t!]
\centering
    \includegraphics[width=\hsize]{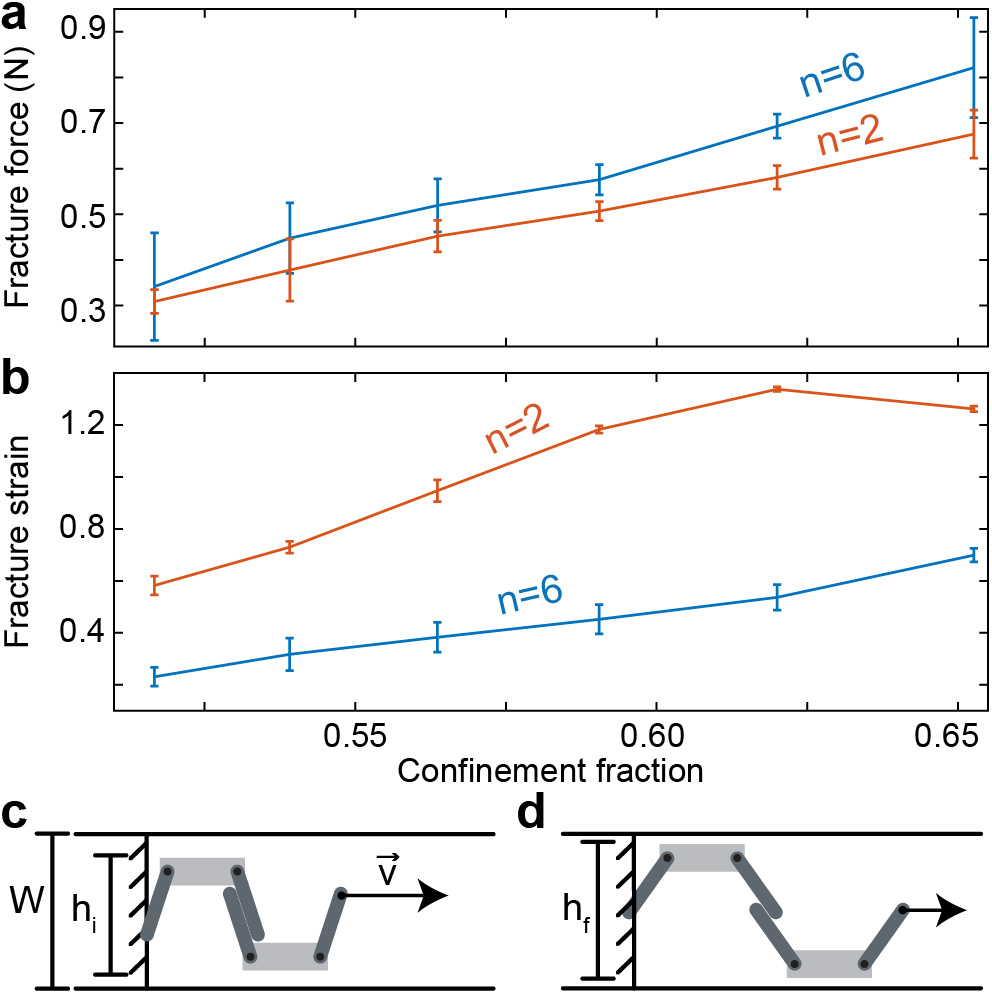}
    \caption[Auxetic behavior in smarticle chain]{\textbf{Auxetic behavior in smarticle chain and dependency on number.} (a) Peak force before fracture (b) peak strain before fracture. (c) Schematic of smarticle chain setup. (d) Configuration after smarticle chain has been strained. $h_f>h_i$.}
    \label{fig:confinement}
\end{figure}

We begin the investigation of some of the elastic properties of our system. With two types of trials, one for each value of $n$, we measure force as a function of strain for the chain. 
There is a small decrease in force at a strain of $\epsilon=0.2$, for $n=2$ and $\epsilon=0.4$ for $n=6$ (see Fig.~\ref{fig:fracture}), the same mechanism causes the decrease in both systems. The decrease in force happens as a result of a sudden and rapid increase in the chain's length. This sudden change in length corresponds to the yield stress, the point at which rearrangement in the chain occurs. Surpassing the yield stress indicates a material has undergone plastic deformation. After reaching this yield stress point during a cyclic strain test, the force as a function of strain will produce a different curve than in cycles before the yield point was reached. As $n$ increases, force also increases but elongation decreases. When arranged in a chain, the PD-controlled servos can be expected to act approximately as a chain of springs for small strains. Therefore, the force should increase linearly with strain distance.

Fracture is defined as the separation of a body into two or more pieces in response to an imposed stress~\cite{Callister2014}. Fracture tests insight into when materials plastically deform and fail. For our second experiment, we measure properties related to fracture in the chain system, namely, the peak force and strain before fracture. Trials were run with $n=[2,6]$ smarticles and the smarticle order was rearranged between each trial. All trials were repeated 5 times for confinement widths between $h/H=[0.52-0.65]$, 
where $h=6.2$~cm is the initial width of the smarticle chain. 
\begin{figure*}[ht!]
\centering
    \includegraphics[width=\hsize]{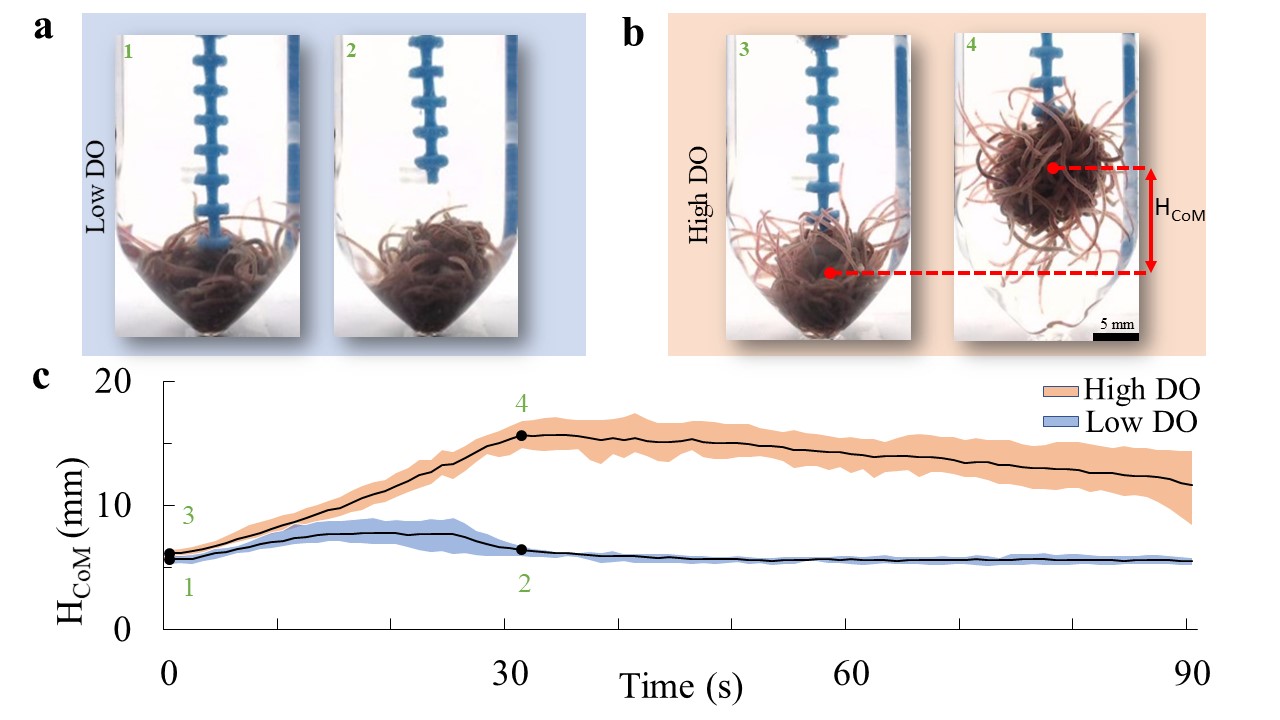}
    \caption[Blob-on-a-stick]{\textbf{Assessing internal mechanical stress} Measuring position of center of mass height (H$_{CoM}$) of worm blob structure as a function of time in two dissolved oxygen environments. Using a 3D-printed serrated endpiece, worm blobs are lifted up by 15 mm using a linear actuator. (a) In low DO ($<$2.0 mg/L), they display less internal mechanical stress by entangling loosely with one another. In high DO ($>$8.0 mg/L) (b), they show increased amounts of internal mechanical stress, enough to be lifted using the endpiece. (c) Position of the H$_{CoM}$ as a function of time in both environments. Shaded region corresponds to the span of $n=5$ trials. Reused from Tuazon, et al. with permission~\cite{2022Tuazon}.}
    \label{fig:blobstick}
\end{figure*}

In the chain system, the fracture mechanism is related to the arm opening angle and friction. When the chain is strained, the barbs' angle is forcefully increased beyond the $\sqcup$-shape or $\alpha_i=90^{\circ}$. As $\alpha_i$ increases with stress, the barb's contact with the adjacent smarticle will begin to slip away from the adjacent smarticle in a direction lateral to the strain direction. The chain fails, or fractures, when the expansion reaches a certain threshold such that the static friction is overcome and adjacent barbs slip lose contact. As the smarticles slip, due to their geometry, they displace outwards. The phenomenon of dilation in the direction lateral to the strain is called auxeticity~\cite{Callister2014}. The width of the chain as it is strained can change, but the system is bounded and has a maximum of $H$ (see Fig.~\ref{fig:confinement}(a)). In trials where the chain width expands to equal the confining width $H$ before it fractures, the maximum force before the fracture is affected. Some of the stress from the chain is offloaded and supported by the walls, effectively reducing the load on the arms as shown in Fig.~\ref{fig:confinement}(a-b). As the confinement increases, ($H$ decreases), the maximal force measured before a fracture will increase. The force increases linearly with the confinement fraction. Moreover, the functional form of an $n=2$ and $n=6$ smarticle chain is qualitatively similar and only the magnitude changes. We find similar results in Fig.~\ref{fig:confinement}(b) for the maximum strain at fracture. As confinement increases, maximum possible strain increases as well. This is consistent with the spring chain approximation of the servo chain.

In our smarticle system, based on ``smarticle cloud'' results ~\cite{Savoie2019}, we hypothesize the existence of gaits which can reliably produce contraction as well as an expansion for a smarticle cloud system. In a contracting case, smarticles at the center of the pile experience confinement. With that as our motivation, we tested how confinement affected fracture. We found that the maximum force before fracture steadily increases with confinement fraction, similarly, we found the maximal strain distance before fracture increases as well. By leveraging future capabilities of smarticle swarms, we could effectively employ other smarticles to enforce the confinement conditions, allowing a chain to exhibit improved tensile strength performance on command. Finally, we compare emergent properties that arises from simulations to physically entangled collective behavior found in biology.

\begin{figure*}[ht!]
\centering
    \includegraphics[width= \hsize]{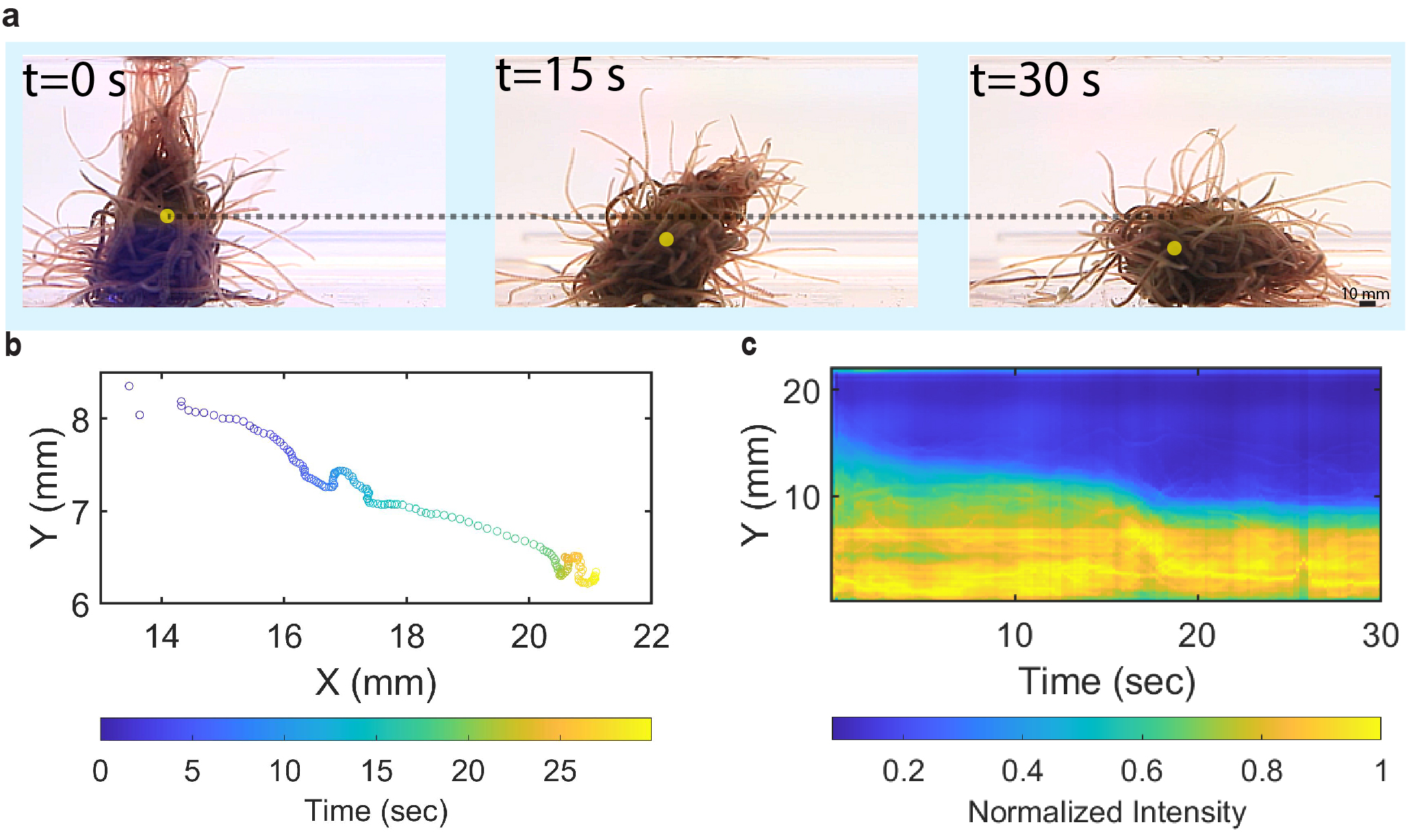}
    \caption[Falling behavior worms.]{\textbf{Tumbling behavior of a worm blob.}(a) Timelapse of a tumbling worm blob ($\sim$2 g) at three instances. At t=0, a short burst of UV 365nm light caused worms to retract their tails. The yellow dot represents the estimated projected center of mass (CoM) (b) CoM XY position as a function of time. (c) Heat map depicting the CoM Y-axis pixel intensities as a function of time. Shaded region corresponds to the span of $n=1$ trials.}
    \label{fig:blobtumble}
\end{figure*}

\subsection{Biological Experiment}
Here, inspired by the results from simulations, we conducted experiments using an analogous entangled system found in nature. Similar to how smarticles display the metallurgical dynamics of annealing, worm blobs can emergently display a few of these behaviors from their entanglement. Of course, we note that in contrast to smarticles, blackworms are highly flexible with aspects ratios >40~\cite{2021Nguyen} that can generate thrust using muscles across its body length~\cite{1989Drewes, 1999Drewes}. Additionally, worms can display a wide range of length as it extends and contracts its body. 

As an analogy to shape-change, worms blobs can display either very high or very low entanglement in steady state environmental parameters. As a comparison to the smarticles simulations, very low levels of entanglement occurs when worms are extended (Fig.~\ref{fig:simAndExpPics}b), similar to when the barbs travel $\pi/2$ outwards (Fig.~\ref{fig:2a}d). However, when transitioning between levels of environmental parameters, worms can reversibly vary their activity levels and change their level of entanglement, an analogy for internal oscillations. 
For example, according to Tuazon, et. al, worm blobs show less activity and less structural exposed surface area in high DO (>8 mg/L) as opposed to hypoxic conditions (<2 mg/L)~\cite{2022Tuazon}. Furthermore, worm blobs in high DO shows higher internal mechanical stress, which was demonstrated by lifting worm blobs using a 3D serrated endpiece (Fig.~\ref{fig:blobstick}a-c) in two extreme levels of DO. Additionally, according to Ozkan-Aydin, et al.~\cite{2021Ozkan-Aydin}, a highly entangled, "solid-like" blob was observed in lower temperatures (T<25$^\circ$C). However, when exposed to higher temperatures T>25$^\circ$C, the blob slowly disentangled into a "fluid-like" aggregate: demonstrating that higher temperature leads to higher internal oscillations followed by full disentanglement. In short, worm blobs in low temperatures or high DO tend to display solid-like structural properties and vice versa.

In a new behavioral observation not found in literature, we observe that blackworms can collectively alter its CoM using the air-water interface as support. As previously discussed, blackworms supplement respiration by lifting their tails up. If the water level is low enough, tails can reach the air-water interface at a 90$^\circ$ and break surface tension. Drewes, et al. deduced that blackworm tails can break surface tension due to hydrophobic material properties found on its skin's surface~\cite{1990Drewes}. We hypothesize that this hydrophobic effect exerts an upward force against buoyancy and gravitational forces which helps to "latch'' and stabilize the tail onto the interface. As worms in a blob reach up and latch their tails on the interface at low DO, its CoM shifts upwards, creating an "inward and grows upward rather than outward'' with a narrow base structure, similar to the shape in the sculpting simulations Fig.~\ref{fig:sculpting}a. We believe that the shift in CoM is due increased internal mechanical stress and higher level of entanglement from a combination of access to higher DO concentrations and from thigmotaxis.

When we stimulated blackworm's negative phototaxis behavior~\cite{2021Ozkan-Aydin, 1990Drewes} using a short burst of UV 395 nm light, individuals immediately retracted their tails, unlatching from the interface and causing the collective to topple over on its side (Fig.~\ref{fig:blobtumble}a-c). This result is analogous to the melting and sculpting experiments for the shape-change trials after the stabilizing structure was removed. Similar to the shape-change results, the structure topples over, due to its unstable conformation from having a narrow base and high CoM, but maintains its overall entanglement.

\section{Conclusions}
We have performed, to our knowledge, the first study of time-dependent material-properties of soft amorphous entangled matter through simulations, robotics and living systems. Despite the seemingly complex interactions which comprise the material properties of a collective, we found entanglement, which encodes much of the material properties, is controllable. Building upon previous studies of entangleable materials, we found that fracture force increased with the degree of entanglement in a simulation. Next, we performed a 2D fracture test analog and found that the peak fracture force increased as a function of confinement due to the auxetic nature of the smarticle chain. Finally, we compared our simulations to the physically-entangled collective behavior blackworms and found that by controlling the ambient DO, worm blobs can display solid-like material properties and tumble from an unstable conformation. 

In the future, we envision an adaptable material capable of maximal or minimal entanglement in response to stimuli (i.e. fire ants' raft formations in the presence of flooding~\cite{Mlot2011,Mlot2012}). This entanglement response should be able to spread through the collective ``organically'', or via local communication, rather than through a global or centralized mechanism. We believe that another major insight from this study is that the choice of how limbs may be activated to alter entanglement is optimizable as a function of energy cost. Due to improvements in the field of rheology over the last century, there has been an explosion in the number of goods with designed material properties based on their purpose: from detergents, to plastics, to hydrophobic clothing~\cite{barnes1989}. We posit that by improving our understanding of the emergent collective dynamics of mutable active particles, it will represent a major step forward in the race towards adaptive, shape-shifting and task-capable soft, super-materials~\cite{Mcevoy2015}.


\section*{Author Contributions}
W.S. designed and carried out the smarticles experiments, analyzed, and interpreted the results. H.T. observed, designed, and carried out the biological experiments, analyzed, and interpreted the results. D.I.G. and M.S.B. reviewed the design and execution of experiments, the data analysis, the interpretations, and the manuscript. All authors contributed to writing the manuscript.

\section*{Conflicts of interest}
There are no conflicts to declare.

\section*{Acknowledgements}
H.T. acknowledges funding support from the NSF graduate research fellowship program (GRFP) and Georgia Tech's President's Fellowship. D.I.G. acknowledges funding support from ARO MURI award (W911NF-19-1-023), NSF Physics of Living Systems Grant (PHY-1205878), and the Dunn Family Professorship award. M.S.B. acknowledges funding support from NIH Grant R35GM142588 and NSF Grants CAREER 1941933 and 1817334.

We thank members of the Bhamla lab and Goldman lab for useful discussions. We especially want thank Akash Vardhan for experimental assistance, Dana Randall, Andr\'{e}a W.~Richa, and Paul Umbanhowar for helpful discussions. Finally, we thank Dr. Prateek Seghal for assisting with the image analysis of the biological experiment. 




\balance

\renewcommand\refname{References}

\bibliography{rsc} 

\providecommand{\noopsort}[1]{}\providecommand{\singleletter}[1]{#1}%
\providecommand*{\mcitethebibliography}{\thebibliography}
\csname @ifundefined\endcsname{endmcitethebibliography}
{\let\endmcitethebibliography\endthebibliography}{}
\begin{mcitethebibliography}{59}
\providecommand*{\natexlab}[1]{#1}
\providecommand*{\mciteSetBstSublistMode}[1]{}
\providecommand*{\mciteSetBstMaxWidthForm}[2]{}
\providecommand*{\mciteBstWouldAddEndPuncttrue}
  {\def\EndOfBibitem{\unskip.}}
\providecommand*{\mciteBstWouldAddEndPunctfalse}
  {\let\EndOfBibitem\relax}
\providecommand*{\mciteSetBstMidEndSepPunct}[3]{}
\providecommand*{\mciteSetBstSublistLabelBeginEnd}[3]{}
\providecommand*{\EndOfBibitem}{}
\mciteSetBstSublistMode{f}
\mciteSetBstMaxWidthForm{subitem}
{(\emph{\alph{mcitesubitemcount}})}
\mciteSetBstSublistLabelBeginEnd{\mcitemaxwidthsubitemform\space}
{\relax}{\relax}

\bibitem[Hu \emph{et~al.}(2016)Hu, Phonekeo, Altshuler, and
  Brochard-Wyart]{2016Hu}
D.~Hu, S.~Phonekeo, E.~Altshuler and F.~Brochard-Wyart, \emph{The European
  Physical Journal Special Topics}, 2016, \textbf{225}, 629--649\relax
\mciteBstWouldAddEndPuncttrue
\mciteSetBstMidEndSepPunct{\mcitedefaultmidpunct}
{\mcitedefaultendpunct}{\mcitedefaultseppunct}\relax
\EndOfBibitem
\bibitem[Reid \emph{et~al.}(2012)Reid, Latty, Dussutour, and Beekman]{2012Reid}
C.~R. Reid, T.~Latty, A.~Dussutour and M.~Beekman, \emph{Proceedings of the
  National Academy of Sciences}, 2012, \textbf{109}, 17490--17494\relax
\mciteBstWouldAddEndPuncttrue
\mciteSetBstMidEndSepPunct{\mcitedefaultmidpunct}
{\mcitedefaultendpunct}{\mcitedefaultseppunct}\relax
\EndOfBibitem
\bibitem[Mlot \emph{et~al.}(2011)Mlot, Tovey, and Hu]{Mlot2011}
N.~J. Mlot, C.~A. Tovey and D.~L. Hu, \emph{Proceedings of the National Academy
  of Sciences}, 2011, \textbf{108}, 7669--7673\relax
\mciteBstWouldAddEndPuncttrue
\mciteSetBstMidEndSepPunct{\mcitedefaultmidpunct}
{\mcitedefaultendpunct}{\mcitedefaultseppunct}\relax
\EndOfBibitem
\bibitem[Ozkan-Aydin \emph{et~al.}(2021)Ozkan-Aydin, Goldman, and
  Bhamla]{2021Ozkan-Aydin}
Y.~Ozkan-Aydin, D.~I. Goldman and M.~S. Bhamla, \emph{Proceedings of the
  National Academy of Sciences}, 2021, \textbf{118}, e2010542118\relax
\mciteBstWouldAddEndPuncttrue
\mciteSetBstMidEndSepPunct{\mcitedefaultmidpunct}
{\mcitedefaultendpunct}{\mcitedefaultseppunct}\relax
\EndOfBibitem
\bibitem[Blackiston \emph{et~al.}(2021)Blackiston, Lederer, Kriegman, Garnier,
  Bongard, and Levin]{2021blackiston}
D.~Blackiston, E.~Lederer, S.~Kriegman, S.~Garnier, J.~Bongard and M.~Levin,
  \emph{Science Robotics}, 2021, \textbf{6}, eabf1571\relax
\mciteBstWouldAddEndPuncttrue
\mciteSetBstMidEndSepPunct{\mcitedefaultmidpunct}
{\mcitedefaultendpunct}{\mcitedefaultseppunct}\relax
\EndOfBibitem
\bibitem[Kriegman \emph{et~al.}(2020)Kriegman, Blackiston, Levin, and
  Bongard]{2020kriegman}
S.~Kriegman, D.~Blackiston, M.~Levin and J.~Bongard, \emph{Proceedings of the
  National Academy of Sciences}, 2020, \textbf{117}, 1853--1859\relax
\mciteBstWouldAddEndPuncttrue
\mciteSetBstMidEndSepPunct{\mcitedefaultmidpunct}
{\mcitedefaultendpunct}{\mcitedefaultseppunct}\relax
\EndOfBibitem
\bibitem[Dumont and Prakash(2014)]{2014Dumont}
S.~Dumont and M.~Prakash, \emph{Molecular Biology of the Cell}, 2014,
  \textbf{25}, 3461--3465\relax
\mciteBstWouldAddEndPuncttrue
\mciteSetBstMidEndSepPunct{\mcitedefaultmidpunct}
{\mcitedefaultendpunct}{\mcitedefaultseppunct}\relax
\EndOfBibitem
\bibitem[Ndlec \emph{et~al.}(1997)Ndlec, Surrey, Maggs, and
  Leibler]{1997Ndelec}
F.~Ndlec, T.~Surrey, A.~C. Maggs and S.~Leibler, \emph{Nature}, 1997,
  \textbf{389}, 305--308\relax
\mciteBstWouldAddEndPuncttrue
\mciteSetBstMidEndSepPunct{\mcitedefaultmidpunct}
{\mcitedefaultendpunct}{\mcitedefaultseppunct}\relax
\EndOfBibitem
\bibitem[Surrey \emph{et~al.}(2001)Surrey, N{\'e}d{\'e}lec, Leibler, and
  Karsenti]{2001Surrey}
T.~Surrey, F.~N{\'e}d{\'e}lec, S.~Leibler and E.~Karsenti, \emph{Science},
  2001, \textbf{292}, 1167--1171\relax
\mciteBstWouldAddEndPuncttrue
\mciteSetBstMidEndSepPunct{\mcitedefaultmidpunct}
{\mcitedefaultendpunct}{\mcitedefaultseppunct}\relax
\EndOfBibitem
\bibitem[Gravish \emph{et~al.}(2012)Gravish, Franklin, Hu, and
  Goldman]{Gravish2012}
N.~Gravish, S.~V. Franklin, D.~L. Hu and D.~I. Goldman, \emph{Phys. Rev.
  Lett.}, 2012, \textbf{108}, 208001\relax
\mciteBstWouldAddEndPuncttrue
\mciteSetBstMidEndSepPunct{\mcitedefaultmidpunct}
{\mcitedefaultendpunct}{\mcitedefaultseppunct}\relax
\EndOfBibitem
\bibitem[Boudet \emph{et~al.}(2021)Boudet, Lintuvuori, Lacouture, Barois,
  Deblais, Xie, Cassagnere, Tregon, Br{\"u}ckner,
  Baret,\emph{et~al.}]{2021boudet}
J.-F. Boudet, J.~Lintuvuori, C.~Lacouture, T.~Barois, A.~Deblais, K.~Xie,
  S.~Cassagnere, B.~Tregon, D.~Br{\"u}ckner, J.-C. Baret \emph{et~al.},
  \emph{Science Robotics}, 2021, \textbf{6}, eabd0272\relax
\mciteBstWouldAddEndPuncttrue
\mciteSetBstMidEndSepPunct{\mcitedefaultmidpunct}
{\mcitedefaultendpunct}{\mcitedefaultseppunct}\relax
\EndOfBibitem
\bibitem[Li \emph{et~al.}(2019)Li, Batra, Brown, Chang, Ranganathan, Hoberman,
  Rus, and Lipson]{2019Li}
S.~Li, R.~Batra, D.~Brown, H.-D. Chang, N.~Ranganathan, C.~Hoberman, D.~Rus and
  H.~Lipson, \emph{Nature}, 2019, \textbf{567}, 361--365\relax
\mciteBstWouldAddEndPuncttrue
\mciteSetBstMidEndSepPunct{\mcitedefaultmidpunct}
{\mcitedefaultendpunct}{\mcitedefaultseppunct}\relax
\EndOfBibitem
\bibitem[Savoie \emph{et~al.}(2019)Savoie, Berrueta, Jackson, Pervan,
  Warkentin, Li, Murphey, Wiesenfeld, and Goldman]{2019Savoie}
W.~Savoie, T.~A. Berrueta, Z.~Jackson, A.~Pervan, R.~Warkentin, S.~Li, T.~D.
  Murphey, K.~Wiesenfeld and D.~I. Goldman, \emph{Science Robotics}, 2019,
  \textbf{4}, eaax4316\relax
\mciteBstWouldAddEndPuncttrue
\mciteSetBstMidEndSepPunct{\mcitedefaultmidpunct}
{\mcitedefaultendpunct}{\mcitedefaultseppunct}\relax
\EndOfBibitem
\bibitem[Anderson(1980)]{1980Anderson}
P.~Anderson, \emph{More is different Science l77, 393-399 (1972); CV Negoita,
  Fuzzy Systems}, 1980\relax
\mciteBstWouldAddEndPuncttrue
\mciteSetBstMidEndSepPunct{\mcitedefaultmidpunct}
{\mcitedefaultendpunct}{\mcitedefaultseppunct}\relax
\EndOfBibitem
\bibitem[Hsueh and Wei(2010)]{Hsueh2010}
C.~H. Hsueh and W.~C.~J. Wei, \emph{Journal of Applied Physics}, 2010,
  \textbf{107}, 024905\relax
\mciteBstWouldAddEndPuncttrue
\mciteSetBstMidEndSepPunct{\mcitedefaultmidpunct}
{\mcitedefaultendpunct}{\mcitedefaultseppunct}\relax
\EndOfBibitem
\bibitem[Brown \emph{et~al.}(2011)Brown, Zhang, Forman, Maynor, Betts,
  DeSimone, and Jaeger]{Brown2011}
E.~Brown, H.~Zhang, N.~A. Forman, B.~W. Maynor, D.~E. Betts, J.~M. DeSimone and
  H.~M. Jaeger, \emph{Phys. Rev. E}, 2011, \textbf{84}, 031408\relax
\mciteBstWouldAddEndPuncttrue
\mciteSetBstMidEndSepPunct{\mcitedefaultmidpunct}
{\mcitedefaultendpunct}{\mcitedefaultseppunct}\relax
\EndOfBibitem
\bibitem[Egres and Wagner(2005)]{Egres2005}
R.~G. Egres and N.~J. Wagner, \emph{Journal of Rheology}, 2005, \textbf{49},
  719--746\relax
\mciteBstWouldAddEndPuncttrue
\mciteSetBstMidEndSepPunct{\mcitedefaultmidpunct}
{\mcitedefaultendpunct}{\mcitedefaultseppunct}\relax
\EndOfBibitem
\bibitem[Kramb and Zukoski(2011)]{Kramb2011}
R.~C. Kramb and C.~F. Zukoski, \emph{Journal of Rheology}, 2011, \textbf{55},
  1069--1084\relax
\mciteBstWouldAddEndPuncttrue
\mciteSetBstMidEndSepPunct{\mcitedefaultmidpunct}
{\mcitedefaultendpunct}{\mcitedefaultseppunct}\relax
\EndOfBibitem
\bibitem[Murphy \emph{et~al.}(2019)Murphy, MacKeith, Roth, and
  Jaeger]{murphy2019}
K.~A. Murphy, A.~K. MacKeith, L.~K. Roth and H.~M. Jaeger, \emph{Granular
  Matter}, 2019, \textbf{21}, 1--6\relax
\mciteBstWouldAddEndPuncttrue
\mciteSetBstMidEndSepPunct{\mcitedefaultmidpunct}
{\mcitedefaultendpunct}{\mcitedefaultseppunct}\relax
\EndOfBibitem
\bibitem[Nelson \emph{et~al.}(2019)Nelson, Schweizer, Rauzan, Nuzzo, Vermant,
  and Ewoldt]{nelson2019}
A.~Z. Nelson, K.~S. Schweizer, B.~M. Rauzan, R.~G. Nuzzo, J.~Vermant and R.~H.
  Ewoldt, \emph{Current Opinion in Solid State and Materials Science}, 2019,
  100758\relax
\mciteBstWouldAddEndPuncttrue
\mciteSetBstMidEndSepPunct{\mcitedefaultmidpunct}
{\mcitedefaultendpunct}{\mcitedefaultseppunct}\relax
\EndOfBibitem
\bibitem[Man \emph{et~al.}(2005)Man, Donev, Stillinger, Sullivan, Russel,
  Heeger, Inati, Torquato, and Chaikin]{Man2005}
W.~Man, A.~Donev, F.~H. Stillinger, M.~T. Sullivan, W.~B. Russel, D.~Heeger,
  S.~Inati, S.~Torquato and P.~M. Chaikin, \emph{Phys. Rev. Lett.}, 2005,
  \textbf{94}, 198001\relax
\mciteBstWouldAddEndPuncttrue
\mciteSetBstMidEndSepPunct{\mcitedefaultmidpunct}
{\mcitedefaultendpunct}{\mcitedefaultseppunct}\relax
\EndOfBibitem
\bibitem[Desmond and Franklin(2006)]{Desmond2006}
K.~Desmond and S.~V. Franklin, \emph{Phys. Rev. E}, 2006, \textbf{73},
  031306\relax
\mciteBstWouldAddEndPuncttrue
\mciteSetBstMidEndSepPunct{\mcitedefaultmidpunct}
{\mcitedefaultendpunct}{\mcitedefaultseppunct}\relax
\EndOfBibitem
\bibitem[Miskin and Jaeger(2014)]{Miskin2014}
M.~Z. Miskin and H.~M. Jaeger, \emph{Soft Matter}, 2014, \textbf{10},
  3708--3715\relax
\mciteBstWouldAddEndPuncttrue
\mciteSetBstMidEndSepPunct{\mcitedefaultmidpunct}
{\mcitedefaultendpunct}{\mcitedefaultseppunct}\relax
\EndOfBibitem
\bibitem[Trepanier and Franklin(2010)]{Trepanier2010}
M.~Trepanier and S.~V. Franklin, \emph{Phys. Rev. E}, 2010, \textbf{82},
  011308\relax
\mciteBstWouldAddEndPuncttrue
\mciteSetBstMidEndSepPunct{\mcitedefaultmidpunct}
{\mcitedefaultendpunct}{\mcitedefaultseppunct}\relax
\EndOfBibitem
\bibitem[Blouwolff and Fraden(2006)]{Blouwolff2006}
J.~Blouwolff and S.~Fraden, \emph{Europhysics Letters ({EPL})}, 2006,
  \textbf{76}, 1095--1101\relax
\mciteBstWouldAddEndPuncttrue
\mciteSetBstMidEndSepPunct{\mcitedefaultmidpunct}
{\mcitedefaultendpunct}{\mcitedefaultseppunct}\relax
\EndOfBibitem
\bibitem[Miskin and Jaeger(2013)]{miskin2013}
M.~Z. Miskin and H.~M. Jaeger, \emph{Nature materials}, 2013, \textbf{12},
  326\relax
\mciteBstWouldAddEndPuncttrue
\mciteSetBstMidEndSepPunct{\mcitedefaultmidpunct}
{\mcitedefaultendpunct}{\mcitedefaultseppunct}\relax
\EndOfBibitem
\bibitem[Gardel \emph{et~al.}(2003)Gardel, Valentine, Crocker, Bausch, and
  Weitz]{Gardel2003}
M.~L. Gardel, M.~T. Valentine, J.~C. Crocker, A.~R. Bausch and D.~A. Weitz,
  \emph{Phys. Rev. Lett.}, 2003, \textbf{91}, 158302\relax
\mciteBstWouldAddEndPuncttrue
\mciteSetBstMidEndSepPunct{\mcitedefaultmidpunct}
{\mcitedefaultendpunct}{\mcitedefaultseppunct}\relax
\EndOfBibitem
\bibitem[Foster \emph{et~al.}(2014)Foster, Mlot, Lin, and Hu]{foster2014}
P.~C. Foster, N.~J. Mlot, A.~Lin and D.~L. Hu, \emph{Journal of Experimental
  Biology}, 2014, \textbf{217}, 2089--2100\relax
\mciteBstWouldAddEndPuncttrue
\mciteSetBstMidEndSepPunct{\mcitedefaultmidpunct}
{\mcitedefaultendpunct}{\mcitedefaultseppunct}\relax
\EndOfBibitem
\bibitem[Mlot \emph{et~al.}(2012)Mlot, Tovey, and Hu]{Mlot2012}
N.~J. Mlot, C.~Tovey and D.~L. Hu, \emph{Communicative \& Integrative Biology},
  2012, \textbf{5}, 590--597\relax
\mciteBstWouldAddEndPuncttrue
\mciteSetBstMidEndSepPunct{\mcitedefaultmidpunct}
{\mcitedefaultendpunct}{\mcitedefaultseppunct}\relax
\EndOfBibitem
\bibitem[McLeish(2008)]{mcleish2008}
T.~McLeish, \emph{Physics today.}, 2008, \textbf{61}, 40--45\relax
\mciteBstWouldAddEndPuncttrue
\mciteSetBstMidEndSepPunct{\mcitedefaultmidpunct}
{\mcitedefaultendpunct}{\mcitedefaultseppunct}\relax
\EndOfBibitem
\bibitem[Lieleg \emph{et~al.}(2009)Lieleg, Schmoller, Cyron, Luan, Wall, and
  Bausch]{lieleg2009}
O.~Lieleg, K.~M. Schmoller, C.~J. Cyron, Y.~Luan, W.~A. Wall and A.~R. Bausch,
  \emph{Soft Matter}, 2009, \textbf{5}, 1796--1803\relax
\mciteBstWouldAddEndPuncttrue
\mciteSetBstMidEndSepPunct{\mcitedefaultmidpunct}
{\mcitedefaultendpunct}{\mcitedefaultseppunct}\relax
\EndOfBibitem
\bibitem[Nguyen \emph{et~al.}(2021)Nguyen, Ozkan-Aydin, Tuazon, Goldman,
  Bhamla, and Peleg]{2021Nguyen}
C.~Nguyen, Y.~Ozkan-Aydin, H.~Tuazon, D.~I. Goldman, M.~S. Bhamla and O.~Peleg,
  \emph{Frontiers in Physics}, 2021, \textbf{9}, 1--12\relax
\mciteBstWouldAddEndPuncttrue
\mciteSetBstMidEndSepPunct{\mcitedefaultmidpunct}
{\mcitedefaultendpunct}{\mcitedefaultseppunct}\relax
\EndOfBibitem
\bibitem[Deblais \emph{et~al.}(2020)Deblais, Maggs, Bonn, and
  Woutersen]{2020aDeblais}
A.~Deblais, A.~Maggs, D.~Bonn and S.~Woutersen, \emph{Physical Review Letters},
  2020, \textbf{124}, 188002\relax
\mciteBstWouldAddEndPuncttrue
\mciteSetBstMidEndSepPunct{\mcitedefaultmidpunct}
{\mcitedefaultendpunct}{\mcitedefaultseppunct}\relax
\EndOfBibitem
\bibitem[Deblais \emph{et~al.}(2020)Deblais, Woutersen, and Bonn]{2020bDeblais}
A.~Deblais, S.~Woutersen and D.~Bonn, \emph{Physical Review Letters}, 2020,
  \textbf{124}, 208006\relax
\mciteBstWouldAddEndPuncttrue
\mciteSetBstMidEndSepPunct{\mcitedefaultmidpunct}
{\mcitedefaultendpunct}{\mcitedefaultseppunct}\relax
\EndOfBibitem
\bibitem[Henein and White(2007)]{henein2007}
C.~M. Henein and T.~White, \emph{Physica A: statistical mechanics and its
  applications}, 2007, \textbf{373}, 694--712\relax
\mciteBstWouldAddEndPuncttrue
\mciteSetBstMidEndSepPunct{\mcitedefaultmidpunct}
{\mcitedefaultendpunct}{\mcitedefaultseppunct}\relax
\EndOfBibitem
\bibitem[Garcimart{\'\i}n \emph{et~al.}(2015)Garcimart{\'\i}n, Pastor, Ferrer,
  Ramos, Mart{\'\i}n-G{\'o}mez, and Zuriguel]{garcimartin2015}
A.~Garcimart{\'\i}n, J.~Pastor, L.~Ferrer, J.~Ramos, C.~Mart{\'\i}n-G{\'o}mez
  and I.~Zuriguel, \emph{Physical Review E}, 2015, \textbf{91}, 022808\relax
\mciteBstWouldAddEndPuncttrue
\mciteSetBstMidEndSepPunct{\mcitedefaultmidpunct}
{\mcitedefaultendpunct}{\mcitedefaultseppunct}\relax
\EndOfBibitem
\bibitem[Zheng \emph{et~al.}(2009)Zheng, Zhong, and Liu]{zheng2009}
X.~Zheng, T.~Zhong and M.~Liu, \emph{Building and Environment}, 2009,
  \textbf{44}, 437--445\relax
\mciteBstWouldAddEndPuncttrue
\mciteSetBstMidEndSepPunct{\mcitedefaultmidpunct}
{\mcitedefaultendpunct}{\mcitedefaultseppunct}\relax
\EndOfBibitem
\bibitem[Michael \emph{et~al.}(2006)Michael, Belta, and Kumar]{michael2006}
N.~Michael, C.~Belta and V.~Kumar, Proceedings 2006 IEEE International
  Conference on Robotics and Automation, 2006. ICRA 2006., 2006, pp.
  964--969\relax
\mciteBstWouldAddEndPuncttrue
\mciteSetBstMidEndSepPunct{\mcitedefaultmidpunct}
{\mcitedefaultendpunct}{\mcitedefaultseppunct}\relax
\EndOfBibitem
\bibitem[Kelley and Ouellette(2013)]{kelley2013}
D.~H. Kelley and N.~T. Ouellette, \emph{Scientific reports}, 2013, \textbf{3},
  1073\relax
\mciteBstWouldAddEndPuncttrue
\mciteSetBstMidEndSepPunct{\mcitedefaultmidpunct}
{\mcitedefaultendpunct}{\mcitedefaultseppunct}\relax
\EndOfBibitem
\bibitem[Savoie \emph{et~al.}(2019)Savoie, Berrueta, Jackson, Pervan,
  Warkentin, Li, Murphey, Wiesenfeld, and Goldman]{Savoie2019}
W.~Savoie, T.~A. Berrueta, Z.~Jackson, A.~Pervan, R.~Warkentin, S.~Li, T.~D.
  Murphey, K.~Wiesenfeld and D.~I. Goldman, \emph{Science Robotics}, 2019,
  \textbf{4}, eaax4316\relax
\mciteBstWouldAddEndPuncttrue
\mciteSetBstMidEndSepPunct{\mcitedefaultmidpunct}
{\mcitedefaultendpunct}{\mcitedefaultseppunct}\relax
\EndOfBibitem
\bibitem[Savoie \emph{et~al.}(2018)Savoie, Cannon, Daymude, Warkentin, Li,
  Richa, Randall, and Goldman]{Savoie2018}
W.~Savoie, S.~Cannon, J.~J. Daymude, R.~Warkentin, S.~Li, A.~W. Richa,
  D.~Randall and D.~I. Goldman, \emph{Artificial Life and Robotics}, 2018,
  \textbf{23}, 459--468\relax
\mciteBstWouldAddEndPuncttrue
\mciteSetBstMidEndSepPunct{\mcitedefaultmidpunct}
{\mcitedefaultendpunct}{\mcitedefaultseppunct}\relax
\EndOfBibitem
\bibitem[Tuazon \emph{et~al.}(2022)Tuazon, Kaufman, Goldman, and
  Bhamla]{2022Tuazon}
H.~Tuazon, E.~Kaufman, D.~I. Goldman and M.~S. Bhamla, \emph{Integrative and
  Comparative Biology}, 2022\relax
\mciteBstWouldAddEndPuncttrue
\mciteSetBstMidEndSepPunct{\mcitedefaultmidpunct}
{\mcitedefaultendpunct}{\mcitedefaultseppunct}\relax
\EndOfBibitem
\bibitem[Purcell(1977)]{Purcell1977}
E.~M. Purcell, \emph{American Journal of Physics}, 1977, \textbf{45},
  3--11\relax
\mciteBstWouldAddEndPuncttrue
\mciteSetBstMidEndSepPunct{\mcitedefaultmidpunct}
{\mcitedefaultendpunct}{\mcitedefaultseppunct}\relax
\EndOfBibitem
\bibitem[Hatton \emph{et~al.}(2013)Hatton, Ding, Choset, and
  Goldman]{hatton2013}
R.~L. Hatton, Y.~Ding, H.~Choset and D.~I. Goldman, \emph{Phys. Rev. Lett.},
  2013, \textbf{110}, 078101\relax
\mciteBstWouldAddEndPuncttrue
\mciteSetBstMidEndSepPunct{\mcitedefaultmidpunct}
{\mcitedefaultendpunct}{\mcitedefaultseppunct}\relax
\EndOfBibitem
\bibitem[Chrono(2019)]{ChronoWebsite}
P.~Chrono, \emph{Chrono: {An Open Source Framework for the Physics-Based
  Simulation of Dynamic Systems}}, \url{http://projectchrono.org}, 2019,
  Accessed: 2019-01-17\relax
\mciteBstWouldAddEndPuncttrue
\mciteSetBstMidEndSepPunct{\mcitedefaultmidpunct}
{\mcitedefaultendpunct}{\mcitedefaultseppunct}\relax
\EndOfBibitem
\bibitem[Tasora \emph{et~al.}(2015)Tasora, Serban, Mazhar, Pazouki, Melanz,
  Fleischmann, Taylor, Sugiyama, and Negrut]{Tasora2015}
A.~Tasora, R.~Serban, H.~Mazhar, A.~Pazouki, D.~Melanz, J.~Fleischmann,
  M.~Taylor, H.~Sugiyama and D.~Negrut, International Conference on High
  Performance Computing in Science and Engineering, 2015, pp. 19--49\relax
\mciteBstWouldAddEndPuncttrue
\mciteSetBstMidEndSepPunct{\mcitedefaultmidpunct}
{\mcitedefaultendpunct}{\mcitedefaultseppunct}\relax
\EndOfBibitem
\bibitem[Timm and Martin(2015)]{2015Timm}
T.~Timm and P.~J. Martin, in \emph{Clitellata: Oligochaeta}, Elsevier, 2015,
  pp. 529--549\relax
\mciteBstWouldAddEndPuncttrue
\mciteSetBstMidEndSepPunct{\mcitedefaultmidpunct}
{\mcitedefaultendpunct}{\mcitedefaultseppunct}\relax
\EndOfBibitem
\bibitem[Drewes(1990)]{1990Drewes}
C.~D. Drewes, \emph{Journal of the Iowa Academy of Science: JIAS}, 1990,
  \textbf{97}, 112--114\relax
\mciteBstWouldAddEndPuncttrue
\mciteSetBstMidEndSepPunct{\mcitedefaultmidpunct}
{\mcitedefaultendpunct}{\mcitedefaultseppunct}\relax
\EndOfBibitem
\bibitem[Callister(2014)]{Callister2014}
W.~Callister, \emph{Materials science and engineering : an introduction},
  Wiley, Hoboken, NJ, 2014\relax
\mciteBstWouldAddEndPuncttrue
\mciteSetBstMidEndSepPunct{\mcitedefaultmidpunct}
{\mcitedefaultendpunct}{\mcitedefaultseppunct}\relax
\EndOfBibitem
\bibitem[Gravish \emph{et~al.}(2014)Gravish, Umbanhowar, and
  Goldman]{Gravish2014}
N.~Gravish, P.~B. Umbanhowar and D.~I. Goldman, \emph{Physical Review E}, 2014,
  \textbf{89}, 042202\relax
\mciteBstWouldAddEndPuncttrue
\mciteSetBstMidEndSepPunct{\mcitedefaultmidpunct}
{\mcitedefaultendpunct}{\mcitedefaultseppunct}\relax
\EndOfBibitem
\bibitem[Agarwal \emph{et~al.}(2019)Agarwal, Senatore, Zhang, Kingsbury,
  Iagnemma, Goldman, and Kamrin]{Agarwal2019}
S.~Agarwal, C.~Senatore, T.~Zhang, M.~Kingsbury, K.~Iagnemma, D.~I. Goldman and
  K.~Kamrin, \emph{Journal of Terramechanics}, 2019, \textbf{85}, 1--14\relax
\mciteBstWouldAddEndPuncttrue
\mciteSetBstMidEndSepPunct{\mcitedefaultmidpunct}
{\mcitedefaultendpunct}{\mcitedefaultseppunct}\relax
\EndOfBibitem
\bibitem[Scott and Kilgour(1969)]{Scott1969}
G.~Scott and D.~Kilgour, \emph{Journal of Physics D: Applied Physics}, 1969,
  \textbf{2}, 863\relax
\mciteBstWouldAddEndPuncttrue
\mciteSetBstMidEndSepPunct{\mcitedefaultmidpunct}
{\mcitedefaultendpunct}{\mcitedefaultseppunct}\relax
\EndOfBibitem
\bibitem[Andreotti \emph{et~al.}(2013)Andreotti, Forterre, and
  Pouliquen]{Andreotti2013}
B.~Andreotti, Y.~Forterre and O.~Pouliquen, \emph{Granular media: between fluid
  and solid}, Cambridge University Press, 2013\relax
\mciteBstWouldAddEndPuncttrue
\mciteSetBstMidEndSepPunct{\mcitedefaultmidpunct}
{\mcitedefaultendpunct}{\mcitedefaultseppunct}\relax
\EndOfBibitem
\bibitem[Onoda and Liniger(1990)]{Onoda1990}
G.~Y. Onoda and E.~G. Liniger, \emph{Phys. Rev. Lett.}, 1990, \textbf{64},
  2727--2730\relax
\mciteBstWouldAddEndPuncttrue
\mciteSetBstMidEndSepPunct{\mcitedefaultmidpunct}
{\mcitedefaultendpunct}{\mcitedefaultseppunct}\relax
\EndOfBibitem
\bibitem[Wells(1991)]{wells1991}
D.~Wells, \emph{The Penguin dictionary of curious and interesting geometry},
  Penguin Mass Market, 1991, vol.~1\relax
\mciteBstWouldAddEndPuncttrue
\mciteSetBstMidEndSepPunct{\mcitedefaultmidpunct}
{\mcitedefaultendpunct}{\mcitedefaultseppunct}\relax
\EndOfBibitem
\bibitem[Drewes and Fourtner(1989)]{1989Drewes}
C.~D. Drewes and C.~R. Fourtner, \emph{The Biological Bulletin}, 1989,
  \textbf{177}, 363--371\relax
\mciteBstWouldAddEndPuncttrue
\mciteSetBstMidEndSepPunct{\mcitedefaultmidpunct}
{\mcitedefaultendpunct}{\mcitedefaultseppunct}\relax
\EndOfBibitem
\bibitem[Drewes(1999)]{1999Drewes}
C.~D. Drewes, \emph{Hydrobiologia}, 1999, \textbf{406}, 263--269\relax
\mciteBstWouldAddEndPuncttrue
\mciteSetBstMidEndSepPunct{\mcitedefaultmidpunct}
{\mcitedefaultendpunct}{\mcitedefaultseppunct}\relax
\EndOfBibitem
\bibitem[Barnes \emph{et~al.}(1989)Barnes, Hutton, and Walters]{barnes1989}
H.~A. Barnes, J.~F. Hutton and K.~Walters, \emph{An introduction to rheology},
  Elsevier, 1989\relax
\mciteBstWouldAddEndPuncttrue
\mciteSetBstMidEndSepPunct{\mcitedefaultmidpunct}
{\mcitedefaultendpunct}{\mcitedefaultseppunct}\relax
\EndOfBibitem
\bibitem[McEvoy and Correll(2015)]{Mcevoy2015}
M.~A. McEvoy and N.~Correll, \emph{Science}, 2015, \textbf{347}, 1261689\relax
\mciteBstWouldAddEndPuncttrue
\mciteSetBstMidEndSepPunct{\mcitedefaultmidpunct}
{\mcitedefaultendpunct}{\mcitedefaultseppunct}\relax
\EndOfBibitem
\end{mcitethebibliography}
\bibliographystyle{rsc} 

\end{document}